# MIMO Wiretap Channels with Arbitrarily Varying Eavesdropper Channel States

Xiang He   Aylin Yener

Wireless Communications and Networking Laboratory

Electrical Engineering Department

The Pennsylvania State University, University Park, PA 16802

*xxh119@psu.edu   yener@ee.psu.edu*

July 20, 2010

## Abstract

In this work, a class of information theoretic secrecy problems is addressed where the eavesdropper channel states are *completely unknown* to the legitimate parties. In particular, MIMO wiretap channel models are considered where the channel of the eavesdropper is *arbitrarily varying* over time. Assuming that the number of antennas of the eavesdropper is limited, the secrecy rate of the MIMO wiretap channel in the sense of *strong secrecy* is derived, and shown to match with the converse in secure degrees of freedom. It is proved that there exists a *universal* coding scheme that secures the confidential message against any sequence of channel states experienced by the eavesdropper. This yields the conclusion that secure communication is possible regardless of the location or channel states of (potentially infinite number of) eavesdroppers. Additionally, it is observed that, the present setting renders the secrecy capacity problems for multi-terminal wiretap-type channels more tractable as compared the case with full or partial knowledge of eavesdropper channel states. To demonstrate this observation, secure degrees of freedom regions are derived for the Gaussian MIMO multiple access wiretap channel (MIMO MAC-WT) and the Gaussian MIMO broadcast wiretap channel (MIMO BC-WT) where the transmitter(s) and the intended receiver(s) have the same number of antennas.

This work is supported in part by the National Science Foundation via Grants CNS-0716325, CCF-0964362, and the DARPA ITMANET Program via Grant W911NF-07-1-0028. This work will be presented in part at Allerton conference on communication, control and computing, September, 2010, and the IEEE Global Telecommunications conference (Globecom), December, 2010.

# I. INTRODUCTION

Information theoretic secrecy dates back to the seminal work by Shannon [1], where it was shown that if the eavesdropper had perfect knowledge of the signals sent by the transmitter and had unbounded computational power, to achieve perfect secrecy, the transmitter and the receiver would have to share a key whose rate equals that of the data.

Wyner, in [2], found that Shannon's result was overly pessimistic, and showed that for the wiretap channel, where the eavesdropper had a noisy observation of the signals sent by the transmitter, a positive rate could be supported for transmitting confidential messages without requiring the communicating parties to share a key. The model was generalized by Csiszár and Körner in [3].

The wiretap channel model in [2]–[4] has inspired considerable effort toward identifying secure communication limits of various channel models, e.g. [5]–[16]. In these works, it is assumed that the transmitter(s) has (have) perfect knowledge of the eavesdropper channel states, which may be difficult to obtain in a practical system, since the eavesdropper is by nature a passive entity. To resolve this issue, recent works attempt to relax this condition by assuming the transmitter only has *partial* knowledge about the channel states of the eavesdropper. Notably, this line of work includes the compound setting, where the eavesdropper channel can only be taken from a *finite* selection [17]–[20], and the fading channel, where the transmitter only knows the distribution of the eavesdropper channel [21]. These each call for different types of codebook design. For example, in [17]–[19], the coding scheme depends on the possible channel gains of the eavesdropper included in the finite set. For the fading setting [21], the duration of communication needs to be able to accommodate a sufficient number of channel uses to ensure that the ergodicity assumption is valid. In addition, the rate of the codebook depends on the fading parameter of the eavesdropper, e.g., the variance of the Rayleigh distribution, and thus needs to be acquired, which may be difficult to do with a passive malicious entity. Given the absence of a robustness analysis toward understanding how sensitive the achievable secrecy rate is to errors in the aforementioned modeling parameters in [17]–[19], [21], it is difficult to ascertain how close these can model a realistic secure system design based on information theoretic guarantees.

The case where the eavesdropper's location is not perfectly known was also considered in the context of network coding. In [22], the eavesdropper is assumed to monitor no more than $K$ edges in a network, while the locations of these edges can be arbitrary. The code designer uses the fact that there are more than $K$ routes connecting the sender and the receiver of the confidential message, while the eavesdropper can not monitor all routes. The simple, yet powerful insight offered by [22] on the merit of utilizing the advantage enjoyed by the legitimate nodes via multiple routes, can be brought into the wireless setting by utilizing multiple antennas. Specifically, if the intended receiver has more antennas than the eavesdropper, then even though the eavesdropper can be anywhere, i.e., experience any channel state, it can not monitor all antennas of the receiver.

Inspired by this observation, in this work, we study the MIMO wiretap channel, where the number of antennas at the eavesdropper is limited. The channel state sequence of the eavesdropper is assumed to be independent from the input and output of the wiretap channel. Conditioned on any given sequence of channel states, we assume that the eavesdropper channel is memoryless. No restriction is placed on the statistics of the channel model observed by the eavesdropper. That is, the eavesdropper's channel is totally unknown to the legitimate parties and can vary arbitrarily over time.

The main contribution of this work is to prove the existence of a *universal* coding scheme that secures the confidential message against any sequence of eavesdropper channel states for the MIMO wiretap setting described above. This means the coding scheme could withstand the presence of infinitely many eavesdroppers as long as they do not collude. The universal nature of the coding scheme is what sets this work apart from the previous work that considered a similar setting with fading [23]. Additionally, unlike [24] which considered the discrete arbitrarily varying wiretap channel, this work considers a Gaussian setting which does not lend itself to a direct extension from its discrete counterpart.

The achievable rates we prove in this work satisfy *strong* secrecy requirements [25]. It is well accepted that weak secrecy is insufficient for practice [25]–[27]. However, it is often argued that strong secrecy can be obtained from weak secrecy through privacy amplification, as shown in [25]. We shall show that, in the setting considered in this paper, this is not the case. Therefore, not only we provide the proof of weak secrecy, but a direct proof of strong secrecy as well, which yields the same achievable rate as its weak counterpart.

The achieved rate derived in this work is *tight* in terms of secure degrees of freedom (s.d.o.f.), which is a high SNR characteristic of the secrecy capacity.

We also extend our results to a MIMO MAC wiretap channel (MIMO MAC-WT) and a MIMO Broadcast wiretap channel (MIMO BC-WT), for the case where legitimate transmitter(s) and receiver(s) have the same number of antennas, and identify their secure degrees of freedom region. It is interesting to note that, in the seemingly simpler setting where the eavesdropper's channel is perfectly known to the legitimate parties, the secure degrees of freedom region for the MIMO MAC-WT channel remains an open problem [19], [28], [29].

The remainder of the paper is organized as follows. The system models are introduced in Section II. In Section III, we state the main results, which are proved in Sections IV and V. Section VI presents a discussion on strong secrecy as well as a detailed comparison to related work. Section VII concludes the paper.

## II. SYSTEM MODELS

### A. The $(N_T, N_R, N_E)$ MIMO Wiretap Channel

The channel from the transmitter to the intended receiver, i.e., *the main channel*, is assumed to be static. Let $A(i)$ denote the value of the signal $A$ during the $i$th channel use. The input and output of the main channel during the $i$th channel use are related as:

$$\mathbf{Y}_{N_R\times 1}(i) = \mathbf{H}_{N_R\times N_T}\mathbf{X}_{N_T\times 1}(i) + \mathbf{Z}_{N_R\times 1}(i) \tag{1}$$

where the subscripts denote the dimension of each term. $\mathbf{H}$ denotes the $N_R \times N_T$ channel matrix with complex entries[1]. It is assumed that $\mathbf{H}$ has full rank. $\mathbf{Z}$ is a $N_R \times 1$ vector representing the additive noise. $\mathbf{Z}$ is composed of independent rotationally invariant complex Gaussian random variables, each with zero mean and unit variance. $\mathbf{X}$ and $\mathbf{Y}$ are the transmitted and received signals respectively.

The channel from the transmitter to the eavesdropper, i.e., *the eavesdropper channel*, is an arbitrarily varying channel. It can be expressed as:

$$\tilde{\mathbf{Y}}_{N_E\times 1}(i) = \tilde{\mathbf{H}}_{N_E\times N_T}(i)\mathbf{X}_{N_T\times 1}(i) \tag{2}$$

where $\tilde{\mathbf{Y}}(i)$ denotes the signals received by the eavesdropper during the $i$th channel use. $\tilde{\mathbf{H}}_{N_E\times N_T}(i)$ is the channel state matrix for the eavesdropper channel during the $i$th channel use. We use $\tilde{\mathbf{H}}^n$ to denote $\tilde{\mathbf{H}}(1),...,\tilde{\mathbf{H}}(n)$. $\tilde{\mathbf{H}}^n$ is *not* known at the legitimate parties. We also assume $\tilde{\mathbf{H}}^n$ is independent from $\mathbf{X}^n$.

---

[1]Since we assume that the main channel is static, $\mathbf{H}$ remains fixed for all channel uses.

Note that we assume the eavesdropper's channel is noiseless. This is obviously a worst case assumption, and if the eavesdropper's signals are corrupted by additive noise, they can always be considered as a degraded version of the signals received by the eavesdropper considered in this work.

Let $W$ denote the confidential message transmitted to the intended receiver, over $n$ channel uses using $\mathbf{X}^n$. In addition, we assume that there is a local random source $M$ which is only known to the transmitter. $\mathbf{X}^n$ is computed by the transmitter from $W$ and $M$ using the following encoding function $f_n$:

$$\mathbf{X}^n = f_n(W, M) \quad (3)$$

Note that $f_n$ does not depend on $\tilde{\mathbf{H}}$, since the transmitter does not know the channel state of the eavesdropper.

We represent $\mathbf{X}^n$ as a $N_T \times n$ matrix, with the average power constraint[2]

$$\lim_{n\to\infty} \frac{1}{n}\text{trace}(\mathbf{X}^n(\mathbf{X}^n)^H) \leq \bar{P}. \quad (4)$$

Let $\hat{W}$ denote the decoder output of the intended receiver from $\mathbf{Y}^n$. Then we require

$$\lim_{n\to\infty} \Pr(W \neq \hat{W}) = 0 \quad (5)$$

for reliable communication. Additionally, we require the following *strong* secrecy constraint [25] to hold for any distribution of $\tilde{\mathbf{H}}^n$.

$$\lim_{n\to\infty} I(W; \tilde{\mathbf{Y}}^n, \tilde{\mathbf{H}}^n) = 0 \quad (6)$$

When designing the encoder $f_n$, we need a uniform bound on $I(W; \tilde{\mathbf{Y}}^n, \tilde{\mathbf{H}}^n)$ over all possible distributions of eavesdropper channel state sequences for a given $n$, which will determine the minimal codeword length required to achieve a certain level of secrecy. Hence, it is important for the convergence in (6) to be *uniform* for all possible statistics of the eavesdropper channel state sequence.

---

[2]trace denotes the sum of the diagonal elements of a square matrix.

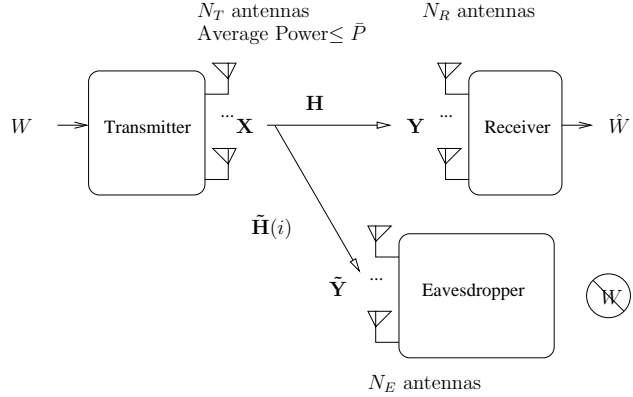

Fig. 1. The MIMO Wiretap Channel.

Since $\tilde{\mathbf{H}}^n$ is independent from $\mathbf{X}^n$, (6) can be written as:

$$\lim_{n\to\infty} I(W; \tilde{\mathbf{Y}}^n | \tilde{\mathbf{H}}^n) = 0 \tag{7}$$

On the other hand, since we do not want the secrecy constraint to rely on the distribution of $\tilde{\mathbf{H}}^n$, we require that for any given sequence $\tilde{\mathbf{h}}^n$,

$$\lim_{n\to\infty} I(W; \tilde{\mathbf{Y}}^n | \tilde{\mathbf{H}}^n = \tilde{\mathbf{h}}^n) = 0. \tag{8}$$

In the sequel, to simplify the notation, we use a scalar $\gamma$ to index the sequence $\tilde{\mathbf{h}}^n$ and use $\tilde{\mathbf{Y}}^n_\gamma$ to represent the random variable $\tilde{\mathbf{Y}}^n$ when $\tilde{\mathbf{H}}^n$ equals the sequence $\tilde{\mathbf{h}}^n$ indexed by $\gamma$. Then (8) can be written as:

$$\lim_{n\to\infty} I(W; \tilde{\mathbf{Y}}^n_\gamma) = 0, \quad \forall \gamma \tag{9}$$

Secrecy rate $R_s$

$$R_s = \lim_{n\to\infty} \frac{1}{n} H(W) \tag{10}$$

for the MIMO wiretap channel is said to be achievable if for each $n$ there exists a *fixed* encoding function $f_n$ as defined by (3), such that for any given the eavesdropper channel $\tilde{\mathbf{H}}$, (5), (9), (4) and (10) holds for $R_s$. The supremum of all possible values for $R_s$ is the *secrecy capacity* of this channel model.

The high SNR behavior of the secrecy rate is characterized by the secure degrees of freedom defined as:

$$\text{s.d.o.f.} = \limsup_{\bar{P}\to\infty} \frac{R_s}{\log_2(\bar{P})} \tag{11}$$

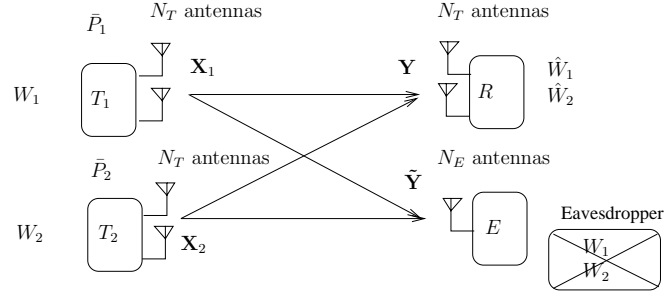

Fig. 2. The $(N_T, N_T, N_T, N_E)$ MIMO MAC wiretap channel where legitimate nodes have $N_T = 2$ antennas each, and the eavesdropper has $N_E = 1$ antenna.

We use the term *the secure degrees of freedom* of a channel to represent the largest possible value of secure degrees of freedom.

## B. The $(N_T, N_T, N_T, N_E)$ MIMO MAC Wiretap Channel

We also consider the MIMO MAC wiretap channel, an example of which is shown in Figure 2. We assume each transmitter has $N_T$ antennas. The receiver also has $N_T$ antennas. The eavesdropper has $N_E$ antennas. During the $i$th channel use, the channel is defined as:

$$\mathbf{Y}(i) = \sum_{k=1}^{2} \mathbf{H}_k \mathbf{X}_k(i) + \mathbf{Z}(i) \tag{12}$$

$$\tilde{\mathbf{Y}}(i) = \sum_{k=1}^{2} \tilde{\mathbf{H}}_k(i) \mathbf{X}_k(i) \tag{13}$$

where $\mathbf{H}_k, k = 1, 2$ and $\tilde{\mathbf{H}}_k(i), k = 1, 2$ are channel matrices. $\mathbf{Z}$ is the additive Gaussian noise observed by the intended receiver, which has the same distribution as the $\mathbf{Z}$ in (1). $\tilde{\mathbf{H}}_k(i), k = 1, 2$ are not known to the legitimate parties. The sequence $\{\tilde{\mathbf{H}}_k(i), k = 1, 2\}$ are independent from the value of $\mathbf{X}_k, k = 1, 2$ over $n$ channel uses. $\mathbf{H}_k, k = 1, 2$ are known by both the legitimate parties *and* the eavesdropper(s).

User $k$ wishes to transmit a confidential message $W_k$ to the receiver over $n$ channel uses, while both messages must be kept confidential from the eavesdropper. We use $\gamma$ to index a specific sequence of $\{\tilde{\mathbf{H}}_k(i), k = 1, 2\}$ over $n$ channel uses and use $\tilde{\mathbf{Y}}_\gamma^n$ to represent the corresponding channel outputs for $\tilde{\mathbf{Y}}^n$. The secrecy constraint is:

$$\lim_{n \to \infty} I\left(W_1, W_2; \tilde{\mathbf{Y}}_\gamma^n\right) = 0, \quad \forall \gamma \tag{14}$$

Again, we stress that the convergence of the limit in (14) must be uniform over all possible sequences of the eavesdropper channel states.

The average power constraints for the two users are given by

$$\lim_{n \to \infty} \frac{1}{n} \text{trace}(\mathbf{X}_k^n (\mathbf{X}_k^n)^H) \leq \bar{P}_k, k = 1, 2 \quad (15)$$

The secrecy rate for $k$, $R_{s,k}$, is defined as

$$R_{s,k} = \lim_{n \to \infty} \frac{1}{n} H(W_k), k = 1, 2 \quad (16)$$

such that $W_k$ can be reliably decoded by the receiver, and (14) and (15) are satisfied.

We define the secure degrees of freedom region similar to [30]: We assume $\bar{P}_k = \bar{P}, k = 1, 2$. The secure degrees of freedom region is defined as:

$$\{(d_1, d_2) : d_k = \limsup_{\bar{P} \to \infty} \frac{R_{s,k}}{\log_2 \bar{P}}, k = 1, 2\} \quad (17)$$

## C. The $(N_T, N_T, N_T, N_E)$ MIMO Broadcast Wiretap Channel

Similarly, we can consider a MIMO Broadcast (BC) wiretap channel shown in Figure 3. We assume that the transmitter has $N_T$ antennas, as well as each receiver. The eavesdropper has $N_E$ antennas. During the $i$th channel use, the channel is defined as:

$$\mathbf{Y}_k(i) = \mathbf{H}_k \mathbf{X}(i) + \mathbf{Z}_k(i), k = 1, 2 \quad (18)$$

$$\tilde{\mathbf{Y}}(i) = \tilde{\mathbf{H}}(i) \mathbf{X}(i) \quad (19)$$

where $\mathbf{H}_k, k = 1, 2$ and $\tilde{\mathbf{H}}(i)$ are the channel matrices. $\mathbf{Z}_k, k = 1, 2$ is the additive Gaussian noise observed by the intended receivers, which has the same distribution as the $\mathbf{Z}$ in (1). $\tilde{\mathbf{H}}(i)$ is unknown to the legitimate parties. $\mathbf{H}_k, k = 1, 2$ are known by both the legitimate parties and the eavesdropper(s).

Each receiver $k$ receives a confidential message $W_k$ from the transmitter over $n$ channel uses, while both messages must be kept confidential from the eavesdropper. The secrecy constraint is:

$$\lim_{n \to \infty} I\left(W_1, W_2; \tilde{\mathbf{Y}}_\gamma^n\right) = 0, \quad \forall \gamma \quad (20)$$

where, as before, $\gamma$ is used to index the eavesdropper channel state sequence. Again the convergence of the limit in (20) must be uniform over all possible sequences of eavesdropper channel states.

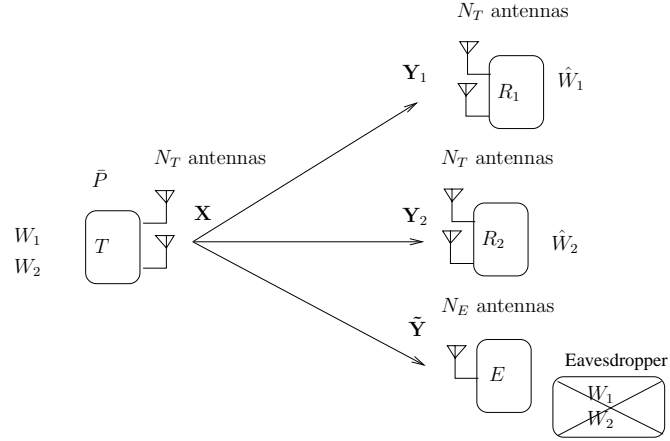

Fig. 3. The $(N_T, N_T, N_T, N_E)$ MIMO BC Wiretap Channel where legitimate nodes have $N_T = 2$ antennas each, and the eavesdropper has $N_E = 1$ antenna.

The average power constraint for the transmitter is given by

$$\lim_{n\to\infty} \frac{1}{n}\text{trace}(\mathbf{X}^n(\mathbf{X}^n)^H) \leq \bar{P} \tag{21}$$

The secrecy rate $R_{s,k}, k = 1, 2$ is defined as in (16). The secure degrees of freedom region $(d_1, d_2)$ is defined as in (17).

## III. MAIN RESULTS

The main results of this work are the following theorems:

*Theorem 1:* Let $N_{T,R} = \min\{N_T, N_R\}$. Let $s_i$ be the $N_{T,R}$ singular values of $\mathbf{H}$. Define $P$ as

$$P = \max\{\bar{P} - N_{T,R}, 0\} \tag{22}$$

Define $C(x)$ as $\frac{1}{2}\log_2(1+x)$. Then any secrecy rate $R_s$ that satisfies

$$0 \leq R_s < \max\left\{\left(\sum_{i=1}^{N_{T,R}} C\left(\frac{s_i^2 P}{(s_i^2+1)N_{T,R}}\right)\right) - N_E C(P), 0\right\} \tag{23}$$

is achievable for the MIMO-wiretap channel described in Section II-A.

Theorem 1 is proved in Section IV.

In Section IV-G, we show that the achievable secrecy rate given by Theorem 1 matches the converse in terms of secure degrees of freedom. Hence we have the following corollary.

*Corollary 1:* If $\mathbf{H}$ has rank $\min\{N_T, N_R\}$, then the s.d.o.f. of the MIMO-wiretap channel described in Section II-A is

$$\max\{\min\{N_T, N_R\} - N_E, 0\} \tag{24}$$

The achievable secrecy rate given by Theorem 1 can be easily extended to the MIMO $(N_T, N_T, N_T, N_E)$ MAC wiretap channel and the MIMO $(N_T, N_T, N_T, N_E)$ BC wiretap channel using time sharing. Let $\alpha$ be the time sharing parameter, and $\bar{\alpha} = 1 - \alpha$. Let $[x]^+$ equal to $x$ if $x \geq 0$ and $0$ if $x < 0$. Define $P_{k,\alpha}, k = 1, 2$ as:

$$P_{k,\alpha} = \left[\frac{\bar{P}_k}{\alpha} - N_T\right]^+, \quad k = 1, 2 \tag{25}$$

Let "co" denote the convex hull operation, and $s_{k,i}, k = 1, 2$ denote the $N_T$ singular values of $\mathbf{H}_k$. Then, for the MIMO $(N_T, N_T, N_T, N_E)$ MAC wiretap channel, the achievable secrecy rate region is given by:

$$\mathrm{co}\bigcup_\alpha \left\{ \begin{array}{l} (R_{s,1}, R_{s,2}): \\ 0 \leq R_{s,1} < \alpha[\sum_{i=1}^{N_T} C\left(\frac{s_{1,i}^2 P_{1,\alpha}}{(s_{1,i}^2+1)N_T}\right) - N_E C\left(P_{1,\alpha}\right)]^+ \\ 0 \leq R_{s,2} < \bar{\alpha}[\sum_{i=1}^{N_T} C\left(\frac{s_{2,i}^2 P_{2,\bar{\alpha}}}{(s_{2,i}^2+1)N_T}\right) - N_E C\left(P_{2,\bar{\alpha}}\right)]^+ \end{array} \right\} \tag{26}$$

For the MIMO $(N_T, N_T, N_T, N_E)$ BC wiretap channel, the achievable secrecy rate region is given by:

$$\mathrm{co}\left\{ \begin{array}{l} (R_{s,1}, R_{s,2}): \quad (0,0), \\ ([\sum_{i=1}^{N_T} C\left(\frac{s_{1,i}^2 P}{(1+s_{1,i}^2)N_T}\right) - N_E C\left(P\right)]^+, 0), \\ (0, [\sum_{i=1}^{N_T} C\left(\frac{s_{2,i}^2 P}{(1+s_{2,i}^2)N_T}\right) - N_E C\left(P\right)]^+) \end{array} \right\} \tag{27}$$

In Section V, we show that these achievable regions match their converse in terms of secure degrees of freedom region. Hence we have the following two theorems:

*Theorem 2:* If $\mathbf{H}_k, k = 1, 2$ has full rank, the secure degrees of freedom region of the MIMO MAC wiretap channel in Figure 2 is given by:

$$d_1 + d_2 \leq \max\{N_T - N_E, 0\} \tag{28}$$

$$d_i \geq 0, \quad i = 1, 2 \tag{29}$$

*Theorem 3:* If $\mathbf{H}_k, k = 1, 2$ has full rank, the s.d.o.f. of the secrecy capacity region of the MIMO BC wiretap channel in Figure 3 is given by:

$$d_1 + d_2 \leq \max\{N_T - N_E, 0\} \tag{30}$$

$$d_i \geq 0, \quad i = 1, 2 \tag{31}$$

*Remark 1:* When the eavesdropper's channel state is fixed and known by the transmitters, the s.d.o.f. region for the MIMO MAC wiretap channel is still an open problem. When the eavesdropper's channel state can take more than one possible value from a finite set and the set is known by the transmitters, the s.d.o.f. region for the MIMO BC wiretap channel is also open. On the other hand if the eavesdropper's channel is arbitrarily varying and is only known by the eavesdropper, and all legitimate nodes have the same number of antennas, the s.d.o.f. capacity region of both problems are found in this paper. □

## IV. THE MIMO WIRETAP CHANNEL

In this section, we present our detailed results on the MIMO wiretap channel. The section is divided into a number of subsections for the reader's convenience. We first present the notation used extensively in the sequel. The next several sections lead to the achievable rate provided by Theorem 1. Section IV-B provides the channel transformation and signaling scheme. Section IV-C presents the codebook construction. Section IV-D presents the achievability proof for the static eavesdropper channel. Building on this proof, Section IV-E presents the achievable secrecy rate for the arbitrarily varying eavesdropper channel. Both of these rates are proved with strong secrecy. The achievable rate with weak secrecy is derived for completeness as well as maintaining consistency with most of current literature in information theoretic secrecy, e.g., [7], [8]. A diagram summarizing the steps leading to the achievability proof along with key methodologies is provided in Figure 4. Finally, in Section IV-G, we present the converse to establish the secure degrees of freedom result in Corollary 1.

### A. Notation

We use $p_W(w)$ to denote the probability mass function (p.m.f.) of a random variable $W$ evaluated at $w$. $f_{\gamma,A}(a)$ denotes the probability density function (p.d.f.) of a random variable $A$ at value $a$ with parameter $\gamma$. $f_{\gamma,A|B}(a|b)$ denotes the conditional p.d.f. of a random variable $A$ conditioned on a random variable $B$ when $A = a, B = b$ with parameter $\gamma$.

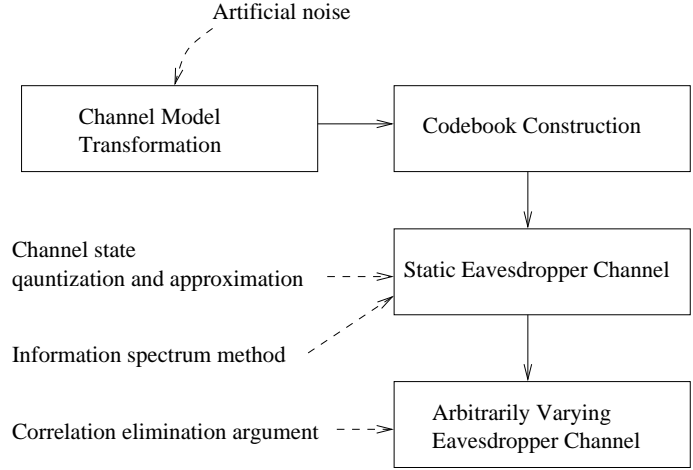

Fig. 4. Organization of the proof for Theorem 1.

For a vector $x^n$, we let $\|x^n\|$ denote its $L_2$-norm. For a matrix $\mathbf{A}$, we let $\|\mathbf{A}\|^2$ denote the sum of the $L_2$-norm squared of all the row vectors of $\mathbf{A}$. $\mathrm{E}_B[A]$ denotes the expectation of $A$ averaged over $B$.

We define

$$\mathbf{A}^n - \mathbf{B}^n\mathbf{C}^n \tag{32}$$

as the row concatenation of the following matrices:

$$\mathbf{A}(i) - \mathbf{B}(i)\mathbf{C}(i), \quad 1 \leq i \leq n \tag{33}$$

and

$$\mathbf{B}^n\mathbf{C}^n \tag{34}$$

as the row concatenation of the following matrices:

$$\mathbf{B}(i)\mathbf{C}(i), \quad 1 \leq i \leq n \tag{35}$$

### B. Channel Model Transformation

Consider a general channel matrix $\mathbf{H}$. We can perform singular value decomposition (SVD) on $\mathbf{H}$ and canceling the right and left unitary matrices of its SVD decomposition at the transmitter and intended receiver respectively. After this cancellation, if $N_T > N_R$, the main channel is equivalent to a MIMO channel whose channel state matrix is $[\mathbf{D}_{N_R \times N_R}, \mathbf{0}_{N_R \times (N_T - N_R)}]$, where $\mathbf{D}$ is a diagonal matrix composed of singular values of $\mathbf{H}$. In this case, we simply use the first $N_R$ antennas at the transmitter only in this equivalent channel when designing the achievable scheme. The remaining $N_T - N_R$ antennas transmit signals of value $0$. If $N_T < N_R$, the main channel is then equivalent to a MIMO link whose channel state matrix is $[\mathbf{D}_{N_T \times N_T}, \mathbf{0}_{N_T \times (N_R - N_T)}]^T$, where $(\ )^T$ means transpose operation. In this case, we simply discard the signals observed at the last $N_R - N_T$ antennas at the receiver in this equivalent channel when designing the achievable scheme. In both cases, from the perspective of designing the achievable scheme, the resulting main channel is equivalent to a MIMO link where the transmitter and the receiver both have $\min\{N_T, N_R\}$ antennas and the channel matrix is diagonal. Thus, without loss of generality, we assume $N_T = N_R$ and $\mathbf{H}$ is a diagonal matrix.

We next observe that in order to prove Theorem 1, we only need to consider $N_E < \min\{N_T, N_R\}$. When $N_E \geq \min\{N_T, N_R\}$, the achievable secrecy rate in Theorem 1 is zero, by virtue of the maximum of the two terms on the right hand side of (23) being zero.

Since $N_E < \min\{N_T, N_R\}$, $\tilde{\mathbf{H}}(i)$ has the following form of SVD decomposition:

$$\tilde{\mathbf{H}}(i) = [\mathbf{I}_{N_E \times N_E}, \mathbf{0}_{N_E \times (N_T - N_E)}] \mathbf{U}(i) \tag{36}$$

where $\mathbf{U}(i)$ is a $N_T \times N_T$ unitary matrix. $\mathbf{I}$ is an identity matrix. This can be achieved by canceling the left unitary matrix of the SVD decomposition of $\tilde{\mathbf{H}}(i)$ and normalizing the singular value at the eavesdropper. Note that the transmitter does not know $\mathbf{U}(i)$.

*Remark 2:* Note that if $\tilde{\mathbf{H}}(i)(\mathbf{U}(i))^{-1}$ has all zero rows, we can alter appropriate entries to be 1s so that the resulting channel matrix has the form in (36). The signals received by the original eavesdropper is always degraded compared to the signals received by the eavesdropper after this modification. Hence, it is sufficient to consider the eavesdroppers with $\tilde{\mathbf{H}}(i)$ in the form given by (36). □

Following [23], we introduce artificial noise at the transmitter. Thus, we have

$$\mathbf{X}(i) = \tilde{\mathbf{X}}(i) + \mathbf{N}(i) \tag{37}$$

where $\mathbf{N}$ is the $N_T \times 1$ artificial noise vector consisting of independent rotationally invariant complex Gaussian random variables with zero mean and unit variance. Coding is over $\tilde{\mathbf{X}}$.

Define $\bar{\mathbf{N}}$ and $\tilde{\mathbf{N}}(i)$ as

$$\bar{\mathbf{N}}(i) = \mathbf{H}\mathbf{N}(i) \tag{38}$$

$$\tilde{\mathbf{N}}(i) = \tilde{\mathbf{H}}(i)\mathbf{N}(i) \tag{39}$$

Viewing $\tilde{\mathbf{X}}$ as the input to the channel, the channel model can be expressed as:

$$\mathbf{Y}(i) = \mathbf{H}\tilde{\mathbf{X}}(i) + \bar{\mathbf{N}}(i) + \mathbf{Z}(i) \tag{40}$$

$$\tilde{\mathbf{Y}}(i) = \tilde{\mathbf{H}}(i)\tilde{\mathbf{X}}(i) + \tilde{\mathbf{N}}(i) \tag{41}$$

From (36) and (39), we observe that $\tilde{\mathbf{N}}$ has zero mean and is Gaussian distributed. The covariance matrix of $\tilde{\mathbf{N}}$ is

$$\mathrm{E}[\tilde{\mathbf{H}}(i)\mathbf{N}(i)(\mathbf{N}(i))^H(\tilde{\mathbf{H}}(i))^H] \tag{42}$$

$$= \tilde{\mathbf{H}}(i)\mathrm{E}[\mathbf{N}(i)(\mathbf{N}(i))^H](\tilde{\mathbf{H}}(i))^H \tag{43}$$

$$= \tilde{\mathbf{H}}(i)(\tilde{\mathbf{H}}(i))^H \tag{44}$$

$$= \mathbf{I}_{N_E \times N_E} \tag{45}$$

### C. Codebook Construction

The *codebook ensemble* we use is constructed as follows:

Recall that $P$ was defined in (22). We choose the input distribution for $\tilde{\mathbf{X}}$, $Q_{\tilde{\mathbf{X}}}(x)$, as rotationally invariant zero mean complex Gaussian with covariance matrix $(\frac{P(1-\varepsilon_P)}{N_T})\mathbf{I}_{N_T \times N_T}$.

The codebook ensemble is composed of the codebooks constructed as described in [31, Section 7.3]. In the context of this work, this means defining the $n$-letter distribution $Q_{\tilde{\mathbf{X}}^n}(x^n)$ as follows: Let $x_i$ denote the $i$th component of $x^n$. $Q_{\tilde{\mathbf{X}}^n}(x^n)$ is given by:

$$Q_{\tilde{\mathbf{X}}^n}(x^n) = \mu_{n,\varepsilon_P}^{-1} \varphi(x^n) \prod_{i=1}^{n} Q_{\tilde{\mathbf{X}}}(x_i) \tag{46}$$

where

$$\varphi(x^n) = \begin{cases} 1, & \text{if } \frac{1}{n}\|x^n\|^2 \leq P \\ 0, & \text{otherwise} \end{cases} \tag{47}$$

$$\mu_{n,\varepsilon_P} = \int \varphi(x^n) \prod_{i=1}^{n} Q_{\tilde{\mathbf{X}}}(x_i) dx^n \tag{48}$$

HE AND YENER, SUBMITTED TO IEEE TRANSACTIONS ON INFORMATION THEORY 15Note that $0 < \mu_{n,\varepsilon_P} < 1$, and for a given $\varepsilon_P > 0$, we have there exists an $\alpha(\varepsilon_P) > 0$, such that [32, (B2)]

$$1 - \mu_{n,\varepsilon_P} \leq e^{-n\alpha(\varepsilon_P)} \tag{49}$$

$$\lim_{\varepsilon_P \to 0} \alpha(\varepsilon_P) = 0 \tag{50}$$

Any codebook in the ensemble is constructed by sampling $2^{nR}$ sequences from the distribution $Q_{\tilde{\mathbf{X}}^n}$ in an independent and identically distributed (i.i.d.) fashion. For the strong secrecy proof, $R$ is chosen as

$$R = I(\tilde{\mathbf{X}}; \mathbf{Y}) - \delta' \tag{51}$$

The mutual information in (51) is evaluated when $\tilde{\mathbf{X}}$ has distribution $Q_{\tilde{\mathbf{X}}}$. $\delta'$ is a positive constant that can be arbitrarily small.

Each time we sample a codeword, we label it with $(i,j)$. Define $N_i$ and $N_j$ as the range of $i$ and $j$. They are given by:

$$N_i = 2^{n(R - I(\tilde{\mathbf{X}}; \tilde{\mathbf{Y}}) - \delta_n)} \tag{52}$$

$$N_j = 2^{n(I(\tilde{\mathbf{X}}; \tilde{\mathbf{Y}}) + \delta_n)} \tag{53}$$

$\{\delta_n\}$ is a positive sequence whose details will be specified later. Again the mutual information in (53) is evaluated when $\tilde{\mathbf{X}}$ has distribution $Q_{\tilde{\mathbf{X}}}$. Note that we drop the subscript $\gamma$ in this expression since the value of the mutual information does not depend on $\gamma$ when $\tilde{\mathbf{X}}$ has the distribution $Q_{\tilde{\mathbf{X}}}$.

The label $i$ takes values from $1, ..., N_i$. $j$ takes values from $1, ..., N_j$. The initial value for $i$ and $j$ are both 1. After we label a codeword, if $j < N_j$, then we increase $j$ by one. Otherwise, we increase $i$ by one and reset $j$ to 1.

Let $\mathcal{C}$ denote a codebook in the codebook ensemble $\{\mathcal{C}\}$. Let $x_{i,j}^n$ denote the codeword in the codebook $\mathcal{C}$ that is labeled with $(i,j)$.

As in [31], for a given codebook $\mathcal{C}$, the intended receiver uses a maximum likelihood decoder: Upon receiving $\mathbf{Y}^n = y^n$, the decoder $\psi_{\mathcal{C}}(y^n)$ is given by

$$\psi_{\mathcal{C}}(y^n) = \arg \max_{i,j: x_{i,j}^n \in \mathcal{C}} \|y^n - \mathbf{H}^n x^n\| \tag{54}$$

The probability of decoding error for each codeword, and the average probability of decoding error for each codebook and the codebook ensembles are defined as:

$$\lambda_{\mathcal{C},i,j} = \Pr\left(\psi_{\mathcal{C}}(\tilde{\mathbf{Y}}^n) \neq (i,j) | \tilde{\mathbf{X}}^n = x_{i,j}^n\right) \quad (55)$$

$$\lambda_{\mathcal{C}} = \frac{1}{N_i N_j} \sum_{i,j} \lambda_{\mathcal{C},i,j} \quad (56)$$

$$\lambda = \mathrm{E}_{\mathcal{C}}\left[\lambda_{\mathcal{C}}\right] \quad (57)$$

We next present the strong secrecy rate when the eavesdropper's channel is static.

### D. Static Eavesdropper Channel

For a given codebook $\mathcal{C}$, the encoder $f_{n,\mathcal{C}}$ used by the transmitter is described as follows: Let the confidential message $W$ be uniformly distributed over the set of $\{i\}$. Given $W = i$, $f_{n,\mathcal{C}}$ selects a codeword from all the codewords with label $i$ in codebook $\mathcal{C}$ according to a uniform distribution. With this encoder, we observe that $(i, j)$ has a uniform distribution.

Let $\tilde{\mathbf{X}}_G^n$ denote $\tilde{\mathbf{X}}^n$ when it is sampled in an i.i.d. fashion from the input distribution $Q_{\tilde{\mathbf{X}}}(x)$ instead of the codebook. Let $\tilde{\mathbf{X}}_T^n$ denote $\tilde{\mathbf{X}}^n$ when it is sampled in an i.i.d. fashion from the $n$-letter truncated Gaussian input distribution $Q_{\tilde{\mathbf{X}}^n}$ instead of the codebook.

Let $\tilde{\mathbf{Y}}_G^n, \tilde{\mathbf{Y}}_T^n, \tilde{\mathbf{Y}}_\mathcal{C}^n$ denote $\tilde{\mathbf{Y}}^n$ when $\tilde{\mathbf{X}}^n$ is $\tilde{\mathbf{X}}_G^n$, $\tilde{\mathbf{X}}_T^n$ or uniformly distributed over the codebook $\mathcal{C}$ respectively.

Let $\gamma$ index (a sequence of) the eavesdropper channel states over $n$ channel uses: $\tilde{\mathbf{H}}_\gamma(1), ..., \tilde{\mathbf{H}}_\gamma(n)$. We use $d_{\gamma,\mathcal{C}}$ to denote the variational distance between two distribution $p_W f_{\gamma, \tilde{\mathbf{Y}}_\mathcal{C}^n}$ and $p_W f_{\gamma, \tilde{\mathbf{Y}}_\mathcal{C}^n | W}$, which is defined as:

$$d_{\gamma,\mathcal{C}} = d\left(p_W f_{\gamma, \tilde{\mathbf{Y}}_\mathcal{C}^n}, p_W f_{\gamma, \tilde{\mathbf{Y}}_\mathcal{C}^n | W}\right) \quad (58)$$

$$= \sum_w \int |p_W(w) f_{\gamma, \tilde{\mathbf{Y}}_\mathcal{C}^n}(z^n) - p_W(w) f_{\gamma, \tilde{\mathbf{Y}}_\mathcal{C}^n | W}(z^n | w)| dz^n \quad (59)$$

$$= \sum_w p_W(w) \int |f_{\gamma, \tilde{\mathbf{Y}}_\mathcal{C}^n}(z^n) - f_{\gamma, \tilde{\mathbf{Y}}_\mathcal{C}^n | W}(z^n | w)| dz^n \quad (60)$$

The proof of achievability we present can be outlined in four steps:

1) As in [27], [33], we first prove for any eavesdropper channel state value indexed by $\gamma$, $d_{\gamma,\mathcal{C}}$ averaged over an ensemble of wiretap codebooks decreases uniformly and exponentially fast with respect to the code length $n$ using the information spectrum method from [34].

2) We then quantize the channel states and construct a finite subset of values of the eavesdropper channel state. We show that for this subset, there must exist a good codebook that retains the property of the codebook ensemble that $d_{\gamma,\mathcal{C}}$ is small.

3) We show that when the eavesdropper channel state is not in the finite subset, the resulting variational distance can be approximated by the variational distance when eavesdropper channel state sequence is in the finite set and hence is also small. This is the approximation argument from [35].

4) Building on 3), we prove that the secrecy constraint is satisfied [33], and hence the codebook secures the message for all possible values of eavesdropper channel states.

We start the proof with the following lemma:

*Lemma 1:* [27, Appendix II, Section D] For a fixed codebook in the ensemble, we have:

$$d_{\gamma,\mathcal{C}} \leq 2\sum_{w} p_W(w) \int |f_{\gamma,\tilde{\mathbf{Y}}_T^n}(z^n) - f_{\gamma,\tilde{\mathbf{Y}}_\mathcal{C}^n|W}(z^n|w)|dz^n \qquad (61)$$

For each integral in the sum in (61), we can write

$$\int |f_{\gamma,\tilde{\mathbf{Y}}_T^n}(z^n) - f_{\gamma,\tilde{\mathbf{Y}}_\mathcal{C}^n|W}(z^n|w)|dz^n \qquad (62)$$

$$\leq \int |f_{\gamma,\tilde{\mathbf{Y}}_G^n}(z^n) - f_{\gamma,\tilde{\mathbf{Y}}_\mathcal{C}^n|W}(z^n|w)|dz^n + \int |f_{\gamma,\tilde{\mathbf{Y}}_G^n}(z^n) - f_{\gamma,\tilde{\mathbf{Y}}_T^n}(z^n)|dz^n \qquad (63)$$

For the second term in (63), we have the following lemma:

*Lemma 2:* For sufficiently large $n$, such that $1/2 > e^{-n\alpha(\varepsilon_P)}$, we have:

$$\int |f_{\gamma,\tilde{\mathbf{Y}}_G^n}(z^n) - f_{\gamma,\tilde{\mathbf{Y}}_T^n}(z^n)|dz^n < 4e^{-n\alpha(\varepsilon_P)} \qquad (64)$$

where $\alpha(\varepsilon_P)$ is the positive exponent defined in (49).

*Proof:* The proof is given in Appendix A. ■

We then can bound the second term in (63) using Lemma 2. We next bound the first term in (63), averaged over $W$.

As in [27, Appendix II, Section D], we can use the symmetry property of the random codebook ensemble and write:

$$\mathrm{E}_\mathcal{C}\left[\sum_{w} p_W(w) \int |f_{\gamma,\tilde{\mathbf{Y}}_G^n}(z^n) - f_{\gamma,\tilde{\mathbf{Y}}_\mathcal{C}^n|W}(z^n|w)|dz^n\right] \qquad (65)$$

$$=\mathrm{E}_\mathcal{C}\left[\int |f_{\gamma,\tilde{\mathbf{Y}}_G^n}(z^n) - f_{\gamma,\tilde{\mathbf{Y}}_\mathcal{C}^n|W}(z^n|1)|dz^n\right]. \qquad (66)$$

See Appendix B for steps to arrive at (66) from (65).

For (66), we have the following lemma:

*Lemma 3:* [34, Lemma 5] For a fixed codebook in the ensemble, we have:

$$\int |f_{\gamma,\tilde{\mathbf{Y}}_G^n}(z^n) - f_{\gamma,\tilde{\mathbf{Y}}_\mathcal{C}^n|W}(z^n|1)|dz^n \leq \frac{2}{\log_2 e}\mu_n + 2\Pr\left[\log_2 \frac{f_{\gamma,\tilde{\mathbf{Y}}_\mathcal{C}^n|W}\left(\tilde{\mathbf{Y}}_\mathcal{C}^n|1\right)}{f_{\gamma,\tilde{\mathbf{Y}}_G^n}\left(\tilde{\mathbf{Y}}_\mathcal{C}^n\right)} > \mu_n\right] \quad (67)$$

Let $f_{\gamma,\tilde{\mathbf{Y}}|\tilde{\mathbf{X}}}$ denote the conditional p.d.f. implied by the channel matrix $\tilde{\mathbf{H}}_\gamma$. Define *information density* [34], $i_{\gamma,\tilde{\mathbf{X}}_G^n\tilde{\mathbf{Y}}_G^n}\left(\tilde{\mathbf{X}}^n,\tilde{\mathbf{Y}}^n\right)$, as :

$$i_{\gamma,\tilde{\mathbf{X}}_G^n\tilde{\mathbf{Y}}_G^n}\left(\tilde{\mathbf{X}}^n,\tilde{\mathbf{Y}}^n\right) = \log_2 \frac{\prod_{i=1}^n f_{\gamma,\tilde{\mathbf{Y}}|\tilde{\mathbf{X}}}\left(\tilde{\mathbf{Y}}_i|\tilde{\mathbf{X}}_i\right)}{f_{\gamma,\tilde{\mathbf{Y}}_G^n}\left(\tilde{\mathbf{Y}}^n\right)} \quad (68)$$

*Remark 3:* Note that the subscript of $i_{\gamma,\tilde{\mathbf{X}}_G^n\tilde{\mathbf{Y}}_G^n}\left(\tilde{\mathbf{X}}^n,\tilde{\mathbf{Y}}^n\right)$ simply indicates the p.d.f.s we use to compute the information spectrum, which are $f_{\gamma,\tilde{\mathbf{Y}}|\tilde{\mathbf{X}}}$ and $f_{\gamma,\tilde{\mathbf{Y}}_G^n}$ in this case. The arguments of $i_{\gamma,\tilde{\mathbf{X}}_G^n\tilde{\mathbf{Y}}_G^n}\left(\tilde{\mathbf{X}}^n,\tilde{\mathbf{Y}}^n\right)$, $\tilde{\mathbf{X}}^n$ and $\tilde{\mathbf{Y}}^n$, can have a different p.d.f. than the one indicated by the subscript of $i_{\gamma,\tilde{\mathbf{X}}_G^n\tilde{\mathbf{Y}}_G^n}\left(\tilde{\mathbf{X}}^n,\tilde{\mathbf{Y}}^n\right)$. □

*Lemma 4:* For any positive $\mu_n$, such that,

$$\tau_n = \left(2^{\mu_n} - \mu_{n,\varepsilon_P}^{-1}\right)/2 > 0 \quad (69)$$

we have:

$$\mathrm{E}_\mathcal{C}\left[\Pr\left[\log_2 \frac{f_{\gamma,\tilde{\mathbf{Y}}_\mathcal{C}^n|W}\left(\tilde{\mathbf{Y}}_\mathcal{C}^n|1\right)}{f_{\gamma,\tilde{\mathbf{Y}}_G^n}\left(\tilde{\mathbf{Y}}_\mathcal{C}^n\right)} > \mu_n\right]\right] \leq \quad (70)$$

$$\mu_{n,\varepsilon_P}^{-1}\{\Pr\left[\frac{1}{n}i_{\gamma,\tilde{\mathbf{X}}_G^n\tilde{\mathbf{Y}}_G^n}\left(\tilde{\mathbf{X}}_G^n,\tilde{\mathbf{Y}}_G^n\right) > I\left(\tilde{\mathbf{X}}_G;\tilde{\mathbf{Y}}_G\right) + \delta_n + \frac{1}{n}\log_2 \tau_n\right] +$$

$$\Pr\left[\frac{1}{n}i_{\gamma,\tilde{\mathbf{X}}_G^n\tilde{\mathbf{Y}}_G^n}\left(\tilde{\mathbf{X}}_G^n,\tilde{\mathbf{Y}}_G^n\right) > I\left(\tilde{\mathbf{X}}_G;\tilde{\mathbf{Y}}_G\right) + \delta_n\right] +$$

$$\frac{1}{\tau_n^2}\left(\Pr\left[\frac{1}{n}i_{\gamma,\tilde{\mathbf{X}}_G^n\tilde{\mathbf{Y}}_G^n}\left(\tilde{\mathbf{X}}_G^n,\tilde{\mathbf{Y}}_G^n\right) > I\left(\tilde{\mathbf{X}}_G;\tilde{\mathbf{Y}}_G\right) + \frac{\delta_n}{2}\right] + 2^{-n\delta_n/2}\right)\} \quad (71)$$

*Remark 4:* The proof of Lemma 4 is adapted from [34, Proof of Theorem 4]. The difference is that, [34, Proof of Theorem 4] would require the expectation to be taken over an ensemble whose codewords are sampled in an i.i.d. fashion from a Gaussian distribution, since (71) is evaluated for this distribution. In Lemma 4, the expectation is over the ensemble whose codewords are sampled from $Q_{\tilde{\mathbf{X}}^n}$, which is close to but not equal to a Gaussian distribution. This difference leads to the term $\mu_{n,\varepsilon_P}^{-1}$ in front of the upper bound given by (71). □

*Proof:* The proof is provided in Appendix C. ∎

*Lemma 5:* For a given $\varepsilon > 0$, if for all $n$, $\delta_n \geq \varepsilon$, then there exists a constant $\alpha'(\varepsilon) > 0$, such that

$$\Pr\left[\frac{1}{n} i_{\gamma, \tilde{\mathbf{X}}_G^n \tilde{\mathbf{Y}}_G^n}\left(\tilde{\mathbf{X}}_G^n, \tilde{\mathbf{Y}}_G^n\right) > I\left(\tilde{\mathbf{X}}_G; \tilde{\mathbf{Y}}_G\right) + \delta_n\right] \leq e^{-n\alpha'(\varepsilon)} \quad (72)$$

*Proof:* The proof is provided in Appendix D. ∎

*Remark 5:* Lemma 5 is termed exponentially information stable [33, Section 2]. □

We next apply Lemma 5 to Lemma 4 and Lemma 3, to bound the first term of (63). Then by using Lemma 2, which bounds the second term on the right hand side of (63), we have the following lemma:

*Lemma 6:* For a given $\varepsilon > 0$, if $\delta_n$ is chosen as:

$$\delta_n \geq \max\{2\varepsilon, \varepsilon + \frac{\alpha(\varepsilon_P)}{2}\log_2 e\} \quad (73)$$

then there exists a constant $c'$ such that for sufficiently large $n$, we have:

$$\mathrm{E}_{\mathcal{C}}\left[2\sum_w p_W(w)\int_{z^n} |f_{\gamma,\tilde{\mathbf{Y}}_G^n}(z^n) - f_{\gamma,\tilde{\mathbf{Y}}_{\mathcal{C}}^n|W}(z^n|w)|dz^n\right] \leq \exp(-c'n) \quad (74)$$

and

$$\mathrm{E}_{\mathcal{C}}\left[d_{\gamma,\mathcal{C}}\right] \leq \mathrm{E}_{\mathcal{C}}\left[2\sum_w p_W(w)\int_{z^n} |f_{\gamma,\tilde{\mathbf{Y}}_T^n}(z^n) - f_{\gamma,\tilde{\mathbf{Y}}_{\mathcal{C}}^n|W}(z^n|w)|dz^n\right] \leq \exp(-c'n) \quad (75)$$

The value of $c'$ depends only on $\varepsilon$ and $\varepsilon_P$. The minimum $n$ for (74) and (75) to hold depends only on $\varepsilon_P$.

*Proof:* The proof is provided in Appendix E. ∎

Note that (75) is the result mentioned in the first step in the proof outline.

We next construct the finite set $S_M$ of quantized eavesdropper channel state values as mentioned in the second step of the proof outline. $S_M$ is defined as follows: If the real and imaginary parts of each element in $M\tilde{\mathbf{H}}_\gamma$ are integers, $\gamma$ is in $S_M$. Note that from (36), $\tilde{\mathbf{H}}\tilde{\mathbf{H}}^H = \mathbf{I}_{N_E \times N_E}$, hence the absolute value of the real and imaginary parts of each element in $\tilde{\mathbf{H}}$ can not exceed 1. Therefore $S_M$ is a finite set with at most $(2M+1)^{2N_T N_E}$ components.

Then from (75), we have:

$$\sum_{\gamma \in S_M} \mathrm{E}_{\mathcal{C}}\left[d_{\gamma,\mathcal{C}}\right] \leq (2M+1)^{2N_T N_E} \exp(-c'n) \tag{76}$$

*Remark 6:* Note that this is the same strategy used in proving the compound channel coding theorem in [35]. Reference [35] considered a discrete memoryless channel which is taken from a potentially infinite set and $S_M$ is constructed by quantizing the channel transition probability matrix. Here, since we are considering the Gaussian channel, doing so will not lead to a finite set. Hence we construct $S_M$ by quantizing the channel gains instead. □

Since the codebook ensemble is constructed as in [31], for some $n_0$, we have, that [31]

$$\lambda \leq 5\exp(-nE(R(\delta'))), \forall n > n_0 \tag{77}$$

where $\lambda$, defined in (57), is the average probability of the decoding error for the codebook ensemble .

Hence, as in [27, Appendix II, Section E], from Markov inequality and (77), we know there must exist one codebook such that

1) The probability of decoding error of the intended receiver vanishes as $n \to \infty$.
2) For each $\gamma \in S_M$, we have

$$d_{\gamma,\mathcal{C}} \leq 2\sum_w p_W(w) \int_{z^n} |f_{\gamma,\tilde{\mathbf{Y}}_G^n}(z^n) - f_{\gamma,\tilde{\mathbf{Y}}_{\mathcal{C}}^n|W}(z^n|w)|dz^n \tag{78}$$

$$\leq 3 \times 2(2M+1)^{2N_T N_E} e^{-c'n} \tag{79}$$

Also, we observe, by our definition of the codebook ensemble that, for this fixed codebook, the average power of each codeword must be smaller or equal to $P$.

This concludes the second step in the proof outline. From here onward, all the discussion is considering on this fixed codebook.

We next evaluate $d_{\gamma,\mathcal{C}}$ when the eavesdropper's channel matrix $\tilde{\mathbf{H}}_\gamma$ is indexed by $\gamma$, $\gamma \notin S_M$.

Let $\mathbf{A}_i$ denote the $i$th row of matrix $\mathbf{A}$. We know there must exist a $\gamma' \in S_M$, such that for the $i$th row of $\Delta_h(k) = \tilde{\mathbf{H}}_\gamma(k) - \tilde{\mathbf{H}}_{\gamma'}(k)$, denoted by $\Delta_{h,i}(k)$, we have:

$$\|\Delta_{h,i}(k)\|^2 < 2N_T/M^2, i=1,...,N_E, k=1,...,n \tag{80}$$

With the notation (34), we define $\Delta_x^n = \Delta_h^n x^n$, $x^n \in \mathcal{C}$. Note that $x^n$ is an $N_T \times n$ matrix and $\mathrm{trace}[x^n(x^n)^H] < nP, \forall x^n \in \mathcal{C}$.

Let $\lambda_{\max}(\mathbf{A})$ be the largest eigenvalue of matrix $\mathbf{A}$. Then for the $i$th row of $\Delta_x^n$, $\Delta_{x,i}^n$, we have:

$$\frac{1}{n}\left\|\Delta_{x,i}^n\right\|^2 \tag{81}$$

$$=\frac{1}{n}\left\|\Delta_{h,i}^n x^n\right\|^2 \tag{82}$$

$$=\frac{1}{n}\sum_{k=1}^n \|\Delta_{h,i}(k)x(k)\|^2 \tag{83}$$

$$\leq \frac{1}{n}\sum_{k=1}^n \lambda_{\max}\left(x(k)(x(k))^H\right)\|\Delta_{h,i}(k)\|^2 \tag{84}$$

$$\leq \frac{1}{n}\sum_{k=1}^n \lambda_{\max}\left(x(k)(x(k))^H\right)\frac{2N_T}{M^2} \tag{85}$$

$$\leq \frac{1}{n}\sum_{k=1}^n \operatorname{trace}\left(x(k)(x(k))^H\right)\frac{2N_T}{M^2} \tag{86}$$

$$\leq \operatorname{trace}\left(\frac{1}{n}\sum_{k=1}^n x(k)(x(k))^H\right)\frac{2N_T}{M^2} \tag{87}$$

$$\leq \frac{2N_T P}{M^2} \tag{88}$$

In (85), we use (80). In (86), we use the fact that the eigenvalues of $\frac{1}{n}x(k)(x(k))^H$ are nonnegative.

It follows then that

$$\frac{1}{n}\|\Delta_x^n\|^2 \leq \frac{2N_T N_E P}{M^2}. \tag{89}$$

For $\varepsilon > 0$, define $r'$ and $r$ as:

$$(r')^2 = \frac{2N_T N_E P}{M^2} \tag{90}$$

$$r = r' + \sqrt{N_E(1+\varepsilon)} \tag{91}$$

With $\tilde{\mathbf{X}}^n$ and $\tilde{\mathbf{Y}}^n$ being the inputs and outputs of the eavesdropper channel with states $\tilde{\mathbf{H}}_\gamma^n$, $\tilde{\mathbf{Y}}^n - \tilde{\mathbf{H}}_\gamma^n \tilde{\mathbf{X}}^n$ is a zero mean rotationally invariant Gaussian distribution whose covariance matrix is an identity matrix. Since from (91), it follows that $r^2 > N_E(1+\varepsilon) > N_E$, there exists a positive $\alpha(\varepsilon)$, such that [32, (B2)]:

$$\Pr\left(\frac{1}{n}\left\|\tilde{\mathbf{Y}}^n - \tilde{\mathbf{H}}_\gamma^n \tilde{\mathbf{X}}^n\right\|^2 \geq r^2 | \tilde{\mathbf{X}}^n = x^n\right) < e^{-n\alpha(\varepsilon)} \tag{92}$$

Note that this bound is uniform regardless of the value of the channel matrix $\tilde{\mathbf{H}}_\gamma^n$.

*Lemma 7:* Define $g(r, r')$ as

$$g(r, r') = r'(2r + r') \tag{93}$$

if we can choose $M$ with respect to $n$ such that

$$ng(r, r') < 1 \tag{94}$$

then there must exist $\gamma' \in S_M$ such that

$$d_{\gamma,\mathcal{C}} \leq 2 \sum_w p_W(w) \int |f_{\gamma, \tilde{\mathbf{Y}}_G^n}(z^n) - f_{\gamma, \tilde{\mathbf{Y}}_{\tilde{\mathcal{C}}}^n | W}(z^n|w)| dz^n + 8e^{-n\alpha(\varepsilon_P)} \tag{95}$$

$$\leq 2 \sum_w p_W(w) \int |f_{\gamma', \tilde{\mathbf{Y}}_G^n}(z^n) - f_{\gamma', \tilde{\mathbf{Y}}_{\tilde{\mathcal{C}}}^n | W}(z^n|w)| dz^n + 12e^{-n\alpha(\varepsilon)} + 4ng(r, r') \tag{96}$$

$$\leq 12(2M + 1)^{2N_T N_E} e^{-c'n} + 12e^{-n\alpha(\varepsilon)} + 4ng(r, r') \tag{97}$$

*Proof:* The proof is provided in Appendix F. ∎

*Remark 7:* The proof of Lemma 7 requires the property that the average energy of each codeword in the codebook does not exceed $P$. This is the reason we need to sample from distribution (46) when we construct the codebook ensemble.

□

*Lemma 8:* There exists a codebook, such that for $c_0 > 0$, we have

$$d_{\gamma,\mathcal{C}} < \exp(-c_0 n), \quad \forall \gamma \tag{98}$$

*Proof:* $g(r, r')$ decrease at the rate of $1/M$. Hence there must exist a positive constant $c_M > 0$, such that $M = \exp(nc_M)$ and both $2(2M + 1)^{2N_T N_E} \exp(-c'n)$ and $ng(r, r')$ decrease exponentially fast to $0$ with respect to $n$. Applying it to Lemma 7, we have Lemma 8. ∎

Let $c_4 = \delta' + \max\{2\varepsilon, \varepsilon + \frac{\alpha(\varepsilon_P)}{2} \log_2 e\}$. From (51), (52) and (73), we observe the codebook rate is given by

$$\lim_{n \to \infty} \frac{1}{n} H(W) \geq I\left(\tilde{\mathbf{X}}; \mathbf{Y}\right) - N_E C\left(P(1 - \varepsilon_P)\right) - c_4 \tag{99}$$

where $C(x) = \log_2(1 + x)$. From (50), we notice that (99) can be made arbitrarily close to

$$I\left(\tilde{\mathbf{X}}; \mathbf{Y}\right) - N_E C(P)) \tag{100}$$

To prove that (100) is an achievable secrecy rate, we need the following lemma from [33], which relates $d_{\gamma,\mathcal{C}}$ to the mutual information $I\left(W;\tilde{\mathbf{Y}}_\gamma^n\right)$:

*Lemma 9:* [33, Lemma 1] Let $|\mathcal{W}|$ be the cardinality of the message set $\mathcal{W}$. Then we have:

$$I\left(W;\tilde{\mathbf{Y}}_\gamma^n\right) \leq d_{\gamma,\mathcal{C}} \log_2 \frac{|\mathcal{W}|}{d_{\gamma,\mathcal{C}}} \tag{101}$$

As shown by Lemma 8, the variational distance $d_{\gamma,\mathcal{C}}$ decreases to $0$ exponentially fast with respect to $n$. Since $W$ is uniformly distributed over $\mathcal{W}$, $\log_2 |\mathcal{W}|$ is $H(W)$. As shown by (99), $H(W)$ increases linearly with $n$. Hence from Lemma 9, $I\left(W;\tilde{\mathbf{Y}}_\gamma^n\right)$ decreases to $0$ exponentially fast with respect to $n$.

This, along with the fact that $W$ is received reliably by the intended receiver and the average power constraint is satisfied by each codeword in the codebook, shows that the rate of the codebook given by (100) is indeed an achievable secrecy rate.

The achieved secrecy rate can then be found by evaluating (100) based on (38) and (40), which leads to (23). This concludes the proof for the static case.

We next extend the result we derived for the static channel to the case where the eavesdropper channel is arbitrarily varying.

### E. Arbitrarily Varying Eavesdropper Channel

When the eavesdropper channel is arbitrarily varying, the second step in the proof for the static case must be modified. This is because, even though the variational distance decreases exponentially fast, the size of the subset $S_M$ of the eavesdropper channel state sequences also increases exponentially fast. In this case, Markov inequality is not sufficient to guarantee the existence of a good codebook and the correlation elimination argument from [36] must be used. The proof outline is as follows:

1) The first step is the same as the static case. We prove for any given sequence of the eavesdropper channel states, the variational distance averaged over an ensemble of wiretap codebooks decreases uniformly and exponentially fast with respect to the code length $n$.

2) Then, for the finite a subset of quantized eavesdropper channel state sequences, we use the correlation elimination argument from [36] to show that there exists a small number of codebooks in the codebook ensemble[3] such that the variational distance averaged over these codebooks is small when the eavesdropper channel state sequence is within the finite set. This is proved by showing that the probability that the variational distance averaged over these codebooks exceeds any given constant is *super*-exponentially small with respect to $n$ for an eavesdropper channel state sequence within the finite subset.

3) The third step is the same as the static case. We show that when the eavesdropper channel state sequence is outside the finite set, the variational distance averaged over this small set of codebooks can be approximated by the variational distance when the eavesdropper channel state sequence is in the finite set and hence is also small. As in the static case proof, a small variational distance implies that the secrecy constraint is satisfied.

4) We then use the small set of codebooks to construct the coding scheme using a two stage transmission scheme introduced in [36].

We next start the proof by defining a normalized version of the variational distance. For a given codebook $\mathcal{C}$, and a given eavesdropper channel state sequence

$$\{\tilde{\mathbf{H}}_\gamma(1), ..., \tilde{\mathbf{H}}_\gamma(n)\}$$

define the normalized variational distance $d'_{\gamma,\mathcal{C}}$ as:

$$d'_{\gamma,\mathcal{C}} = \frac{1}{2} \sum_w p_W(w) \int_{z^n} |f_{\gamma,\tilde{\mathbf{Y}}_G^n}(z^n) - f_{\gamma,\tilde{\mathbf{Y}}_{\mathcal{C}}^n|W}(z^n|w)| dz^n \qquad (102)$$

Clearly, we have

$$0 \leq d'_{\gamma,\mathcal{C}} \leq 1 \qquad (103)$$

Also, from (95) in Lemma 7, we have:

$$d_{\gamma,\mathcal{C}} \leq 4 d'_{\gamma,\mathcal{C}} + 8 e^{-n\alpha(\varepsilon_P)} \qquad (104)$$

We then use Lemma 6 to bound $d'_{\gamma,\mathcal{C}}$. Note that Lemma 6 still holds when the eavesdropper channel is arbitrarily varying.

From (74) in Lemma 6, there must exist a constant $c'$, which only depends on $\varepsilon$ and $\varepsilon_P$ such that

$$\mathrm{E}_{\mathcal{C}}\left[d'_{\gamma,\mathcal{C}}\right] \leq \exp(-c'n) \qquad (105)$$

---

[3]In our case, $K = e^{\varepsilon' n}$ codebooks, where $\varepsilon'$ is a positive constant that can be made arbitrarily small.

Applying (105) to (104), we complete the first step in the proof outline.

We next use the correlation elimination argument from [36] and consider $K$ codebooks, each generated as described in Section IV-C. Denote the $k$th codebook with $\mathcal{C}_k$. $\mathcal{C}_k$ is a random variable, and so is $\frac{1}{K}\sum_{k=1}^{K} d'_{\gamma,\mathcal{C}_k}$. Since for different $k$, $\mathcal{C}_k$ are i.i.d., $d'_{\gamma,\mathcal{C}_k}$ are also i.i.d.. These facts, along with (103), mean that the derivation in [36, (4.1)-(4.5)] can be applied here. In particular, the $j$, $T_j$, $\varepsilon$ and $R$ in [36] corresponds to $k$, $d'_{\gamma,\mathcal{C}_k}$, $c'$ and $K$ here respectively. Consider a positive sequence $\{\epsilon_n\}$. Reference [36, (4.1)-(4.5)] shows that if (105) holds, then for $\alpha' > 0$ and for $n$ such that:

$$1 + e^{\alpha'} e^{-c'n} \leq e^{\epsilon_n} \tag{106}$$

we have:

$$\Pr\left(\frac{1}{K}\sum_{k=1}^{K} d'_{\gamma,\mathcal{C}_k} \geq \epsilon_n\right) \leq e^{-(\alpha'-1)K\epsilon_n}. \tag{107}$$

Let $\alpha' = 2$. Then we have

$$\Pr\left(\frac{1}{K}\sum_{k=1}^{K} d'_{\gamma,\mathcal{C}_k} \geq \epsilon_n\right) \leq e^{-\epsilon_n K}. \tag{108}$$

Define the following finite set $S_M$ of $\gamma$: If the real and imaginary parts of each element in $M\tilde{\mathbf{H}}_\gamma(i)$ are integers for all $i = 1,...,n$, then $\gamma \in S_M$. Note that from (36), $\tilde{\mathbf{H}}(i)(\tilde{\mathbf{H}}(i))^H = \mathbf{I}_{N_E \times N_E}$, hence the absolute value of the real and imaginary parts of each element in $\tilde{\mathbf{H}}(i)$ can not exceed 1. Therefore $S_M$ is a finite set with at most $(2M+1)^{2N_T N_E n}$ components.

Let $|S_M|$ denote the size of the set $S_M$. Then we have:

$$\Pr\left(\frac{1}{K}\sum_{k=1}^{K} d'_{\gamma,\mathcal{C}_k} < \epsilon_n, \forall \gamma \in S_M\right) \tag{109}$$

$$\geq 1 - |S_M|\Pr\left(\frac{1}{K}\sum_{k=1}^{K} d'_{\gamma,\mathcal{C}_k} > \epsilon_n\right) \tag{110}$$

$$= 1 - |S_M|e^{-\epsilon_n K} \tag{111}$$

When $\gamma \notin S_M$, from (96) in Lemma 7, we have that if

$$ng(r,r') < 1 \tag{112}$$

then there must exist $\gamma' \in S_M$, such that

$$d'_{\gamma,\mathcal{C}} \leq d'_{\gamma',\mathcal{C}} + e^{-n\alpha(\varepsilon)} + ng(r,r'). \tag{113}$$

Recall that $r$ and $r'$ were defined in (91) and (90) respectively. It follows that we have:

$$\frac{1}{K}\sum_{k=1}^{K} d'_{\gamma,\mathcal{C}_k} \leq \frac{1}{K}\sum_{k=1}^{K} d'_{\gamma',\mathcal{C}_k} + e^{-n\alpha(\varepsilon)} + ng(r,r'). \tag{114}$$

From Markov inequality and (77), we have:

$$\Pr\left(\lambda_{\mathcal{C}_k} > 5nKe^{-Np(R(\delta'))}\right) \leq \frac{1}{nK} \tag{115}$$

Therefore:

$$\Pr\left(\exists k : \lambda_{\mathcal{C}_k} > 5nKe^{-nE(R(\delta'))}\right) \tag{116}$$

$$\leq \sum_{k=1}^{K} \Pr\left(\lambda_{\mathcal{C}_k} > 5nKe^{-nE(R(\delta'))}\right) \tag{117}$$

$$\leq \frac{1}{n} \tag{118}$$

Or equivalently

$$\Pr\left(\lambda_{\mathcal{C}_k} \leq 5nKe^{-nE(R(\delta'))}, k=1,...,K\right) \geq 1 - \frac{1}{n}. \tag{119}$$

We next choose $\epsilon_n$, the number of codebooks $K$ and the variable $M$, which controls the size of the set $|S_M|$ carefully such that for sufficiently large $n$,

1) (106) is satisfied.
2) $\frac{1}{K}\sum_{k=1}^{K} d'_{\gamma,\mathcal{C}_k}$ in (109) vanishes with high probability for $\gamma \in S_M$.
3) $\frac{1}{K}\sum_{k=1}^{K} d'_{\gamma,\mathcal{C}_k}$ on the left hand side of (114) vanishes for $\gamma \notin S_M$. Note that in order to use the bound (114) to prove this result, (112) must be satisfied and the right hand side of (114) must vanish as well, which relies on 2).
4) $\lambda_{\mathcal{C}_k}, k=1,...,K$ in (119) vanishes with high probability.

Satisfying these conditions leads to the claim that there exists $K$ good codebooks.

A proper choice for $\epsilon_n$, $K$ and $M$ is as follows: Recall that $\alpha(\varepsilon)$ was defined in (92). $\alpha(\varepsilon_P)$ was defined in (49) and (50). For a positive constant $\varepsilon'$ such that

$$\varepsilon' < c' \tag{120}$$

$$\varepsilon' < \alpha(\varepsilon) \tag{121}$$

$$\varepsilon' < \alpha(\varepsilon_P) \tag{122}$$

$$2\varepsilon' < E(R(\delta')) \tag{123}$$

$\epsilon_n$, $K$ and $M$ are chosen to be:

$$\epsilon_n = e^{-n\varepsilon'} \tag{124}$$

$$K = e^{2\varepsilon'n} \tag{125}$$

$$M = e^{2\varepsilon'n} \tag{126}$$

$c'$ in (120) was given in (105).

We first check if these choices satisfy (106). We observe that, since $\epsilon_n > 0$, the right hand side of (106) is lower bounded as:

$$e^{\epsilon_n} \geq 1 + \epsilon_n \tag{127}$$

which, due to (124), equals:

$$1 + e^{-\varepsilon'n} \tag{128}$$

Due to (120), we find (128) is greater than the left hand side of (106) for sufficiently large n such that

$$1 + e^2 e^{-c'n} < 1 + e^{-\varepsilon'n} \tag{129}$$

Hence, (106) is satisfied.

Next we observe from (124) and (125) that

$$e^{-\epsilon_n K} = e^{-e^{-n\varepsilon'} e^{2n\varepsilon'}} = e^{-e^{n\varepsilon'}} \tag{130}$$

We also observe that, due to (126), for sufficiently large $n$,

$$2M + 1 \leq e^{4\varepsilon'n} \tag{131}$$

Hence,

$$|S_M| = (2M+1)^{2N_T N_E n} < e^{8N_T N_E \varepsilon' n^2} \tag{132}$$

Therefore, from (130) and (132), we have:

$$\lim_{n \to \infty} |S_M| e^{-\epsilon_n K} = 0 \tag{133}$$

This means (111) will converge to 1 when $n$ goes to $\infty$. Since $\epsilon_n$ is shown by (124) to converge to 0 when $n$ goes to $\infty$, we observe, from (109)-(111), that $\frac{1}{K}\sum_{k=1}^{K} d'_{\gamma,\mathcal{C}_k}$ in (109) vanishes with high probability.

We next examine (114). We observe from (90), (91) and (93) that $g(r,r')$ decreases at the rate of $1/M$ which, according to (126), equals $e^{-2\varepsilon'n}$. Hence for sufficiently large $n$, we have

$$ng(r,r') < e^{-1.5\varepsilon'n} < e^{-\varepsilon'n} = \epsilon_n \tag{134}$$

holds and (112) is satisfied.

Also, due to (121), we have

$$e^{-n\alpha(\varepsilon)} < e^{-n\varepsilon'} = \epsilon_n \tag{135}$$

If, for the $\gamma'$ in (114),

$$\frac{1}{K}\sum_{k=1}^{K} d'_{\gamma',\mathcal{C}_k} < \epsilon_n \tag{136}$$

then, from (134), (135) and (114), we have

$$\frac{1}{K}\sum_{k=1}^{K} d'_{\gamma,\mathcal{C}_k} < 3\epsilon_n \tag{137}$$

Hence, from (109)-(111), (133) and (136)-(137), we observe there must exist $K$ codebooks, where $K$ is given by (125), such that for any $k$ and $\gamma$,

$$\lambda_{\mathcal{C}_k} \leq 5nKe^{-nE(R(\delta'))} \tag{138}$$

$$\frac{1}{K}\sum_{k=1}^{K} d'_{\gamma,\mathcal{C}_k} < 3\epsilon_n \tag{139}$$

$\epsilon_n$ is given by (124).

We next check if our choice of $K$ leads to a vanishing $\lambda_{\mathcal{C}_k}$. By applying (125) to the right hand side of (138), we find it equals:

$$5ne^{-n(E(R(\delta'))-2\varepsilon')} \tag{140}$$

Due to (123), we find that the right hand side of (138) converges to 0 when $n$ goes to $\infty$. This means:

$$\lim_{n\to\infty} \lambda_{\mathcal{C}_k} = 0, \quad \forall k \tag{141}$$

We next express (139) in terms of $d_{\gamma,\mathcal{C}_k}$. Due to (122), we have, from (104), for sufficiently large $n$:

$$12\epsilon_n = 12e^{-n\varepsilon'} > 8e^{-n\alpha(\varepsilon_P)} \tag{142}$$

Hence (139) implies

$$\frac{1}{K}\sum_{k=1}^{K} d_{\gamma,\mathcal{C}_k} < 24\epsilon_n, \quad \forall \gamma \tag{143}$$

(143) concludes the second and third steps in the proof outline.

We next use the $K$ codebooks to construct the coding scheme. Let the confidential message $W$ be uniformly distributed over the set of $\{1, ..., N_i\}$. The encoder $f_n$ used by the transmitter is described as follows:

1) In the first stage, the transmitter chooses the value for an integer $K'$ from $\{1,...,K\}$ according to a uniform distribution. Given $W = i$, $f_n$ outputs the label $(i,j)$ computed by $f_{n,\mathcal{C}_{K'}}$.

2) In the second stage, $K'$ is transmitted to the intended receiver using a good channel codebook for the main channel.

The decoder of the intended receiver first decode $K'$, then decode the confidential message using $\psi_{n,\mathcal{C}_{K'}}$.

Let $\hat{K}'$ be the result decoded by the intended receiver for $K'$. Then

$$\Pr\left(W \neq \hat{W}\right) \tag{144}$$

$$\leq \Pr\left(K' \neq \hat{K}'\right) + \Pr\left(W \neq \hat{W} | K' = \hat{K}'\right) \tag{145}$$

$$= \Pr\left(K' \neq \hat{K}'\right) + \frac{1}{K}\sum_{k=1}^{K} \lambda_{\mathcal{C}_k} \tag{146}$$

Since

$$\lim_{n\to\infty} \Pr\left(K' \neq \hat{K}'\right) = 0 \tag{147}$$

and (141) holds, we have $\lim_{n\to\infty} \Pr\left(W \neq \hat{W}\right) = 0$.

The variational distance for this coding scheme, $d_\gamma$, is given by

$$d_\gamma = d\left(p_W p_{K'} f_{\gamma,\tilde{Y}^n_{\mathcal{C}_{K'}}}, p_W p_{K'} f_{\gamma,\tilde{Y}^n_{\mathcal{C}_{K'}}|W}\right) \tag{148}$$

$$= \sum_{k,w} p_W(w) \, p_{K'}(k) \int |f_{\gamma, \tilde{Y}^n_{\mathcal{C}_k}}(z^n) - f_{\gamma, \tilde{Y}^n_{\mathcal{C}_k}|W}(z^n|w)| dz^n \qquad (149)$$

$$= \frac{1}{K} \sum_{k=1}^{K} d_{\gamma, \mathcal{C}_k} \qquad (150)$$

From (143) and (124), we observe that (150) decreases at the speed of $e^{-\varepsilon' n}$. Then from Lemma 9, we have:

$$\lim_{n \to \infty} I\left(W; K', \tilde{\mathbf{Y}}^n_\gamma\right) = 0, \quad \forall \gamma \qquad (151)$$

The convergence is uniform over all possible eavesdropper channel matrix sequences and the limit decreases *exponentially fast* with respect to $n$. Let $n'$ denote the total number of channel uses. Then the second stage takes $n_2$ channel uses with $n_2$ given by:

$$n_2 = \frac{1}{R} \log_2 K = \frac{2\varepsilon' \log_2 e}{R_0} n \qquad (152)$$

where $R_0 > 0$ is the rate of the conventional channel codebook $\mathcal{C}_0$. The first stage takes $n$ channel uses. Therefore

$$n' = n + n_2 = \left(\frac{2\varepsilon' \log_2 e}{R_0} + 1\right) n \qquad (153)$$

Define $c(\varepsilon')$ as

$$c(\varepsilon') = \frac{2\varepsilon' \log_2 e}{R_0} + 1 \qquad (154)$$

which can be made arbitrarily close to 1 by making $\varepsilon'$ small.

Let $\tilde{\mathbf{Y}}^{n_2}_\gamma$ denote the signals received by the eavesdropper during the second stage. Then

$$\lim_{n' \to \infty} I\left(W; \tilde{\mathbf{Y}}^{n'}_\gamma\right) \qquad (155)$$

$$= \lim_{n' \to \infty} I\left(W; \tilde{\mathbf{Y}}^n_\gamma, \tilde{\mathbf{Y}}^{n_2}_\gamma\right) \qquad (156)$$

$$\leq \lim_{n' \to \infty} I\left(W; K', \tilde{\mathbf{Y}}^n_\gamma\right) \qquad (157)$$

$$= \lim_{n \to \infty} I\left(W; K', \tilde{\mathbf{Y}}^n_\gamma\right) \qquad (158)$$

$$= 0, \quad \forall \gamma \qquad (159)$$

Let $c_4 = \delta' + \max\{2\varepsilon, \varepsilon + \frac{\alpha(\varepsilon_P)}{2} \log_2 e\}$. The secrecy rate is then given by:

$$\lim_{n' \to \infty} \frac{1}{n'} H(W) \geq (I\left(\tilde{\mathbf{X}}; \mathbf{Y}\right) - N_E C\left(P(1 - \varepsilon_P)\right) - c_4) c(\varepsilon') \qquad (160)$$

From (50), we notice (160) can be made arbitrarily close to

$$I\left(\tilde{\mathbf{X}}; \mathbf{Y}\right) - N_E C\left(P\right) \quad (161)$$

Therefore, the same secrecy rate as given in (23) is achievable even when the eavesdropper channel is arbitrarily varying.

*Remark 8:* In order to use the correlation elimination argument from [36], we made three modifications to its proof:

1) Instead of using average error probability as in [36], we use the normalized variational distance defined in (102).
2) In [36], only $K = n^2$ codebooks are used. Here, in order to use Lemma 9 to bound the mutual information with the variational distance, we use $K = e^{\varepsilon' n}$ codebooks.
3) In [36], the index of the codebook used at the transmitter, i.e., $K'$, needs to be reliably communicated to the receiver over an arbitrarily varying channel. In this work, $K'$ is transmitted using a good channel codebook for the main channel which is static. On the other hand, $K'$ may or may not be reliably received over the arbitrarily varying eavesdropper channel. We simply assume $K'$ is revealed to the eavesdropper in order to compute the lower bound on the achievable secrecy rate which is our goal.

□

## F. Achievable Rate with Weak Secrecy with a Static Eavesdropper Channel

The weak secrecy constraint is expressed as

$$\lim_{n\to\infty} \frac{1}{n} I(W; \tilde{\mathbf{Y}}_\gamma^n) = 0, \quad \forall \gamma. \quad (162)$$

The proof of the weak secrecy rate differs from that for the strong secrecy rate in Section IV-D. In particular, instead of using the information spectrum method from [34], we use an argument that is similar to [17]. This means that we prove the existence of a codebooks composed of a number of sub-codebooks with the following property: Each sub-codebook is a good channel code for the eavesdropper channel[4]. The entire codebook is a good channel code for the main channel. These properties allow us to bound the equivocation as in [17].

---

[4] That is to say that if the eavesdropper knows the transmitted signals are restricted to a certain sub-codebook, it can decode the transmitted confidential message reliably.

We start the proof by describing the codebook ensemble. Any codebook in the ensemble is constructed by sampling $2^{nR}$ sequences in an i.i.d. fashion from the distribution $Q_{\tilde{\mathbf{X}}^n}$ with $\varepsilon_P = 0$, where $Q_{\tilde{\mathbf{X}}^n}$ was given by (46).

For weak secrecy, $R$ is chosen as

$$R = I(\tilde{\mathbf{X}}; \mathbf{Y}) - 2\delta_n \tag{163}$$

$\{\delta_n\}$, which will be specified later, is a positive sequence that converges to $0$ when $n$ goes to $\infty$.

Each time we sample a codeword, we label it with $(i,j)$, $i \in \{1, ..., N_i\}$, $j \in \{1, ..., N_j\}$, where $N_i$ and $N_j$ are given by

$$N_i = 2^{n(R - I(\tilde{\mathbf{X}}; \tilde{\mathbf{Y}}) + \delta_n)} \tag{164}$$

$$N_j = 2^{n(I(\tilde{\mathbf{X}}; \tilde{\mathbf{Y}}) - \delta_n)} \tag{165}$$

Note the difference of the choice of $N_j$ in (165) from that used in the strong secrecy proof, i.e., (53).

Labeling is done as it was in the strong secrecy proof: Initially $i = j = 1$. After we label a codeword, if $j < N_j$, then $j$ is incremented by $1$. Otherwise, $i$ is incremented by $1$ and $j$ is reset to $1$.

The intended receiver uses the same decoding rule as in (54).

We also define a fictitious decoder $\phi_\gamma$ used by the eavesdropper whose channel state matrix is $\tilde{\mathbf{H}}_\gamma$. The decoder computes $j$ given $i = i_0$ and $\tilde{\mathbf{Y}}^n = \tilde{y}^n$, and is a maximum likelihood decoder:

$$\phi_\gamma(\tilde{y}^n) = \arg \max_{j: x^n_{i_0,j} \in \mathcal{C}} \|\tilde{y}^n - \tilde{\mathbf{H}}_\gamma x^n_{i_0,j}\| \tag{166}$$

With the decoders defined, we can then, as in [17, Appendix A], define the following probabilities of error when the codeword $\mathbf{X}^n_{i,j}$ is transmitted:

We use $\eta_{\gamma,j|i}$ to denote the error probability for the eavesdropper to reliably decode $j$ given $i$ and $\tilde{\mathbf{Y}}^n$ when the channel state matrix is $\tilde{\mathbf{H}}_\gamma$.

Let the distribution of $i, j$ as $p_{i,j}$. Then we can define the average probability of decoding error, $\eta_\gamma$, as

$$\eta_\gamma = \sum_{i,j} p_{i,j}(i,j) \eta_{\gamma,j|i} \tag{167}$$

In (167), $p_{i,j}$ is determined by the encoder $f_n$ used by the transmitter, which we shall specify next. Let the confidential message $W$ be uniformly distributed over the set of $\{i\}$. Given $W = i$, $f_n$ selects a codeword from all the codewords with label $i$ according to a uniform distribution. With this encoder, we note that $p_{i,j}$ is uniform, and therefore

$$H(\tilde{\mathbf{X}}^n) = nR \tag{168}$$

For the intended receiver, we simply follow the definitions of the probabilities of decoding error in (55)-(57).

We then use the following facts:

*Lemma 10:* For the codebook ensemble described above, $\mathrm{E}_\mathcal{C}[\eta_\gamma]$ is the same for all $\gamma$.

*Proof:* The proof is provided in Appendix G. ∎

Then, as in the proof of the strong secrecy rate, we quantize the channel gains of the eavesdropper channel and construct the same finite set $S_M$ we used in the strong secrecy proof.

From [31, (7.3.22)], we know that there exists an error exponent $E(R) > 0$ such that, for some $n_0$,

$$\lambda = \mathrm{E}_\mathcal{C}[\lambda_\mathcal{C}] \leq 5\exp(-nE(R)), \forall n > n_0. \tag{169}$$

By the same argument, for an eavesdropper whose channel matrix is $\tilde{\mathbf{H}}_\gamma$ indexed by $\gamma \in S_M$, we know there exists an error exponent $E'_\gamma(\tilde{R}) > 0$ such that for some $n_0$,

$$\mathrm{E}_\mathcal{C}[\eta_\gamma] \leq 5\exp(-nE'_\gamma(\tilde{R})), \forall n > n_0 \tag{170}$$

where $\tilde{R} = I(\tilde{\mathbf{X}}, \tilde{\mathbf{Y}}) - \delta_n$.

*Remark 9:* The constant factor of 5 in (169)-(170) and the requirement $n > n_0$ is used to bound the term $(1/\mu)^{1+\rho}$ in [31, (7.3.22)], since $\rho \leq 1$ and $\lim_{n\to\infty} \mu = 1/2$. Hence $n_0$ is not a function of $\gamma$. □

Note that by Lemma 10, $\mathrm{E}_\mathcal{C}[\eta_\gamma]$ is not a function of $\gamma$. Hence, if an error exponent holds for a certain $\gamma$, it holds for all $\gamma$. Therefore, we can omit the subscript $\gamma$ in the error exponent and rewrite it as $E'(\tilde{R})$.

Recall that $2\delta_n = I(\tilde{\mathbf{X}}, \mathbf{Y}) - R$. Hence, we can rewrite both $E(R)$ and $E'(\tilde{R})$ as a function of $\delta_n$. From [31], $E(R)$ and $E'(R)$ have the following property: For $R > 0$ and $\tilde{R} > 0$:

1) $E(R)$ and $E'(\tilde{R})$ are both positive.
2) Both $E(R)$ and $E'(\tilde{R})$ are monotonically decreasing functions of $\delta_n$.

3) If $\delta_n \to 0$, then both $E(R)$ and $E'(\tilde{R})$ converge to $0$.

Let $\bar{E}(\delta_n) = \max\{E(R), E'(\tilde{R})\}$. Then $\bar{E}(\delta_n)$ also has the three properties listed above. From the linearity of expectation, for sufficiently large $n$ that does not depend on $\gamma$, we can write:

$$\mathrm{E}_{\mathcal{C}}\left[\lambda_{\mathcal{C}} + \sum_{\gamma \in S_M} \eta_\gamma\right] \leq 5((2M+1)^{2N_T N_E} + 1)e^{-\bar{E}(\delta_n)n} \tag{171}$$

This means there must exist one codebook in the ensemble such that

$$\lambda_{\mathcal{C}} \leq 5((2M+1)^{2N_T N_E} + 1)e^{-\bar{E}(\delta_n)n} \tag{172}$$

$$\eta_\gamma \leq 5((2M+1)^{2N_T N_E} + 1)e^{-\bar{E}(\delta_n)n} \tag{173}$$

We next consider the equivocation for one eavesdropper whose channel matrix is in the set $S_M$, with this codebook:

$$H\left(W|\tilde{\mathbf{Y}}_\gamma^n\right) \tag{174}$$

$$= H\left(W|\tilde{\mathbf{Y}}_\gamma^n\right) - H\left(W|\tilde{\mathbf{Y}}_\gamma^n, \tilde{\mathbf{X}}^n\right) \tag{175}$$

$$= I\left(W; \tilde{\mathbf{X}}^n|\tilde{\mathbf{Y}}_\gamma^n\right) \tag{176}$$

$$= H\left(\tilde{\mathbf{X}}^n|\tilde{\mathbf{Y}}_\gamma^n\right) - H\left(\tilde{\mathbf{X}}^n|W, \tilde{\mathbf{Y}}_\gamma^n\right) \tag{177}$$

Applying Fano's inequality [37] to the second term in (177), we find it is upper bounded by:

$$H\left(\tilde{\mathbf{X}}^n|\tilde{\mathbf{Y}}_\gamma^n\right) - 1 - \eta_\gamma nR \tag{178}$$

Applying (173) to (178), we find it is upper bounded by:

$$H\left(\tilde{\mathbf{X}}^n|\tilde{\mathbf{Y}}_\gamma^n\right) - 1 - 5nR\left((2M+1)^{2N_T N_E} + 1\right)e^{-\bar{E}(\delta_n)n} \tag{179}$$

Define $\varepsilon_n$ as:

$$\varepsilon_n = \frac{1}{n} + 5R\left((2M+1)^{2N_T N_E} + 1\right)e^{-\bar{E}(\delta_n)n} \tag{180}$$

Then, we can rewrite (179) as:

$$H\left(\tilde{\mathbf{X}}^n\right) - I\left(\tilde{\mathbf{X}}^n; \tilde{\mathbf{Y}}_\gamma^n\right) - n\varepsilon_n \tag{181}$$

$$\geq H\left(\tilde{\mathbf{X}}^n\right) - h\left(\tilde{\mathbf{Y}}_\gamma^n\right) + h\left(\tilde{\mathbf{Y}}_\gamma^n|\tilde{\mathbf{X}}^n\right) - n\varepsilon_n \tag{182}$$

$$= nR - h\left(\tilde{\mathbf{Y}}_\gamma^n\right) + h\left(\tilde{\mathbf{Y}}_\gamma^n|\tilde{\mathbf{X}}^n\right) - n\varepsilon_n \tag{183}$$

where (183) follows by applying (168).

To proceed, we need the following lemma:

*Lemma 11:*

$$h\left(\tilde{\mathbf{Y}}_\gamma^n\right) \leq N_E n \log \pi e (P+1), \quad \forall \gamma \tag{184}$$

*Proof:* The proof is given in Appendix H. ∎

Using Lemma 11 and (168) and the fact that $h\left(\tilde{\mathbf{Y}}_\gamma^n | \tilde{\mathbf{X}}^n\right) = nN_E \log \pi e$, we find (183) is upper bounded by:

$$nR - nN_E \log_2(1+P) - n\varepsilon_n \tag{185}$$

$$= n\left(I\left(\tilde{\mathbf{X}}; \mathbf{Y}\right) - N_E C(P)\right) - n(2\delta_n + \varepsilon_n). \tag{186}$$

Hence

$$H\left(W | \tilde{\mathbf{Y}}_\gamma^n\right) \geq n\left(I\left(\tilde{\mathbf{X}}; \mathbf{Y}\right) - N_E C(P)\right) - n(2\delta_n + \varepsilon_n). \tag{187}$$

On the other hand, $H(W)$ is given by:

$$H(W) = n\left(I\left(\tilde{\mathbf{X}}; \mathbf{Y}\right) - I\left(\tilde{\mathbf{X}}; \tilde{\mathbf{Y}}_\gamma\right)\right) - n\delta_n \tag{188}$$

$$= n\left(I\left(\tilde{\mathbf{X}}; \mathbf{Y}\right) - N_E C(P)\right) - n\delta_n. \tag{189}$$

Therefore we have

$$I\left(W; \tilde{\mathbf{Y}}_\gamma^n\right) \leq n(\delta_n + \varepsilon_n). \tag{190}$$

We next derive an upper bound on $I\left(W; \tilde{\mathbf{Y}}_\gamma^n\right)$ when the eavesdropper's channel matrix is not in $S_M$.

Let $r^2 = 2N_E$ and $(r')^2 = \frac{2N_T N_E P}{M^2}$. Define $\alpha$ as in (92) with $r^2 = 2N_E$. Let $g_{r,r'} = r'(2r+r')$.

Let $\tilde{\mathbf{H}}_\gamma$ denote the channel matrix of the eavesdropper. From our construction of $S_M$, we know there must exist an $\tilde{\mathbf{H}}_{\gamma'}$ indexed by $\gamma' \in S_M$ such that (80) holds.

We let the eavesdropper with channel matrix $\tilde{\mathbf{H}}_\gamma$ use the same "fictitious" decoder we designed for the eavesdropper with channel matrix $\tilde{\mathbf{H}}'_{\gamma'}$. $\eta_\gamma$ be the corresponding probability of decoding error with this decoder. Then we have the following lemma:

*Lemma 12:*

$$\eta_\gamma \leq e^{-n\alpha} + 5((2M+1)^{2N_T N_E} + 1)e^{-(\bar{E}(\delta_n) - g(r,r'))n} \quad (191)$$

*Proof:* The proof is provided in Appendix I. ∎

We next repeat the equivocation computation in (174)-(190). This yields:

$$I\left(W; \tilde{\mathbf{Y}}_\gamma^n\right) \leq n\left(\delta_n + \varepsilon_n'\right) \quad (192)$$

where

$$\varepsilon_n' = \frac{1}{n} + R\left(e^{-n\alpha} + 5((2M+1)^{2N_T N_E} + 1)\right)e^{-(\bar{E}(\delta_n) - g(r,r'))n} \quad (193)$$

The last step of the achievability proof requires choosing $\delta_n$ carefully with respect to $n$, such that $\varepsilon_n$ and $\varepsilon_n'$ goes to 0 as $n$ goes to $\infty$. This can be done by choosing $\bar{E}(\delta_n)$ properly as follows:

1) $\bar{E}(\delta_n)$ decreases to 0 at the rate of $n^{-1/2}$, which ensures $n\bar{E}(\delta_n) \to \infty$ as $n \to \infty$.
2) $M$ increases at the rate of $n$, hence $g(r,r')$ decreases at the rate of $n^{-1}$. Therefore $n(\bar{E}(\delta_n) - g(r,r')) \to \infty$ at the rate of $\exp(-c_1\sqrt{n})$ for $c_1 > 0$, as $n \to \infty$.

Since $\bar{E}(\delta_n)$ is a monotonically decreasing function of $\delta_n$, this means $\delta_n$ converges to 0 as $n \to \infty$. We also observe in this case both $\varepsilon_n$ and $\varepsilon_n'$ converge uniformly to 0 as $n \to \infty$ for *any eavesdropper channel matrix*.

In summary, we have shown that for any eavesdropper, with the same codebook $\mathcal{C}$, we always have

$$\lim_{n \to \infty} \frac{1}{n} I\left(W; \tilde{\mathbf{Y}}_\gamma^n\right) = 0 \quad (194)$$

The convergence is uniform over all possible values of the eavesdropper channel states.

The reliability requirement (5) is fulfilled by (172).

The average power constraint (4) is guaranteed by the way the codebook ensemble is constructed.

*Remark 10:* The secrecy rate found (189) is identical to the rate we derived with strong secrecy requirement in (100). □

*G. Converse for Corollary 1*

In this section, we establish the result in Corollary 1, by providing the converse for the high SNR characterization of the secrecy rate found in (161).

Since $\tilde{\mathbf{H}}$ can be arbitrary, when $N_E \geq N_T$, we can choose $\tilde{\mathbf{H}}$ as $[\mathbf{I}_{N_T \times N_T}, \mathbf{0}_{N_T \times (N_E - N_T)}]^T$. The eavesdropper in this case has perfect knowledge of the transmitted signal. Clearly, the secrecy capacity is $0$.

We next consider the case when $N_E < N_T$. We use $X_i^j$ to denote the $i$th to the $j$th component in a vector $\mathbf{X}$. The secrecy rate is upper bounded by [4]:

$$I\left(\mathbf{X}; \mathbf{Y} | \tilde{\mathbf{Y}}\right) \tag{195}$$

When $N_T \geq N_R$, we assume $\mathbf{H} = [\mathbf{D}_{N_R \times N_R}, \mathbf{0}_{N_R \times (N_T - N_R)}]$ for a diagonal matrix $\mathbf{D}_{N_R \times N_R}$[5]. Since $\tilde{\mathbf{H}}$ is arbitrary, we choose $\tilde{\mathbf{H}}$ as $[\mathbf{I}_{N_E \times N_E}, \mathbf{0}_{N_E \times (N_T - N_E)}]$. Then (195) equals:

$$I\left(\mathbf{X}; \mathbf{D}_{N_R \times N_R} X_1^{N_R} + \mathbf{Z} | X_1^{N_E}\right) \tag{196}$$

$$= I\left(X_1^{N_R}, X_{N_R+1}^{N_T}; \mathbf{D}_{N_R \times N_R} X_1^{N_R} + \mathbf{Z} | X_1^{N_E}\right) \tag{197}$$

$$= I\left(X_1^{N_R}; \mathbf{D}_{N_R \times N_R} X_1^{N_R} + \mathbf{Z} | X_1^{N_E}\right) + I\left(X_{N_R+1}^{N_T}; \mathbf{D}_{N_R \times N_R} X_1^{N_R} + \mathbf{Z} | X_1^{N_E}, X_1^{N_R}\right) \tag{198}$$

$$= I\left(X_1^{N_R}; \mathbf{D}_{N_R \times N_R} X_1^{N_R} + \mathbf{Z} | X_1^{N_E}\right) \tag{199}$$

When $N_T < N_R$, we assume $\mathbf{H} = [\mathbf{D}_{N_T \times N_T}, \mathbf{0}_{N_T \times (N_R - N_T)}]^T$ for a diagonal matrix $\mathbf{D}_{N_T \times N_T}$. We use the same $\tilde{\mathbf{H}}$ as we did in the previous case. Then (195) equals:

$$I\left(\mathbf{X}; \mathbf{D}_{N_T \times N_T} X_1^{N_T} + Z_1^{N_T}, Z_{N_T+1}^{N_R} | X_1^{N_E}\right) \tag{200}$$

$$= I\left(\mathbf{X}; \mathbf{D}_{N_T \times N_T} X_1^{N_T} + Z_1^{N_T} | X_1^{N_E}\right) + I\left(\mathbf{X}; Z_{N_T+1}^{N_R} | \mathbf{D}_{N_T \times N_T} X_1^{N_T} + Z_1^{N_T}, X_1^{N_E}\right) \tag{201}$$

$$= I\left(\mathbf{X}; \mathbf{D}_{N_T \times N_T} X_1^{N_T} + Z_1^{N_T} | X_1^{N_E}\right) \tag{202}$$

Define $N_m = \min\{N_T, N_R\}$. Then, in both cases, (195) can be written as:

$$I\left(X_1^{N_m}; \mathbf{D}_{N_m \times N_m} X_1^{N_m} + Z_1^{N_m} | X_1^{N_E}\right) \tag{203}$$

$$= I\left(X_1^{N_m}; Y_1^{N_m} | X_1^{N_E}\right) \tag{204}$$

which equals:

$$I(X_{N_E+1}^{N_m}; Y_1^{N_m} | X_1^{N_E}) \tag{205}$$

---

[5]Else, we can perform SVD on $\mathbf{H}$ and transform it into this form.

$$=h(Y_{N_E+1}^{N_m}|X_1^{N_E}) + h(Y_1^{N_E}|X_1^{N_E}, Y_{N_E+1}^{N_m}) - h(Y_1^{N_m}|X_1^{N_m}) \tag{206}$$

$$\leq h(Y_{N_E+1}^{N_m}) + h(Y_1^{N_E}|X_1^{N_E}, Y_{N_E+1}^{N_m}) - h(Y_1^{N_m}|X_1^{N_m}) \tag{207}$$

$$\leq h(Y_{N_E+1}^{N_m}) + h(Y_1^{N_E}|X_1^{N_E}) - h(Y_1^{N_m}|X_1^{N_m}) \tag{208}$$

$$=h(Y_{N_E+1}^{N_m}) + h(Z_1^{N_E}|X_1^{N_E}) - h(Z_1^{N_m}|X_1^{N_m}) \tag{209}$$

$$=h(Y_{N_E+1}^{N_m}) + h(Z_1^{N_E}) - h(Z_1^{N_m}) \tag{210}$$

$$=h(Y_{N_E+1}^{N_m}) - h(Z_{N_E+1}^{N_m}) \tag{211}$$

$$=h(Y_{N_E+1}^{N_m}) - h(Y_{N_E+1}^{N_m}|X_{N_E+1}^{N_m}) \tag{212}$$

$$=I(X_{N_E+1}^{N_m}; Y_{N_E+1}^{N_m}) \tag{213}$$

Since we assume $\mathbf{H}$ of the original MIMO wiretap channel has a full rank, $\mathbf{D}_{N_m \times N_m}$ also has full rank. Hence the elements on the diagonal line of $\mathbf{D}$ are all positive. This means equation (213) increases at a rate of $O((\min\{N_T, N_R\} - N_E)C(\bar{P}))$. Hence we have proved the converse of Corollary 1.

## V. EXTENSION TO THE MIMO MAC WIRETAP CHANNEL AND THE MIMO BROADCAST WIRETAP CHANNEL

In this Section, we extend our results to secure communication in multiuser MIMO channels with arbitrarily varying eavesdropper channel states. In particular, Theorems 2 and 3 are proved next.

We first prove Theorem 2 for the MIMO MAC-WT. We observe that the secrecy rate region in (26) is achieved by time sharing, i.e., letting user 1 transmit during $\alpha$ fraction of the channel uses alone, and user 2 for the remaining channel uses. The region then readily follows from Theorem 1. The achievability of the s.d.o.f. region follows from the achievable secrecy rate region (26), by letting $P_1 = P_2 = P \to \infty$.

For the converse, we simply combine the two transmitters. The channel then becomes a single-user MIMO wiretap channel, where the transmitter has $2N_T$ antennas, the receiver has $N_T$ antennas, and the eavesdropper has $N_E$ antenna. $d_1 + d_2 \leq \max\{N_T - N_E, 0\}$ then follows from the converse of Corollary 1.

We next prove Theorem 3 for the MIMO BC-WT. Again we first obtain an achievable secrecy rate region by time sharing. The secrecy rates achieved by receivers 1 and 2 are found to be (27) as a consequence of Theorem 1. The achieved s.d.o.f. region then follows from the achievable secrecy rate region (27), by letting $P_1 = P_2 = P \to \infty$.

For the converse, we simply combine the two receivers. The channel then becomes a single-user MIMO wiretap channel, where the transmitter has $N_T$ antennas, the receiver has $2N_T$ antennas, and the eavesdropper has $N_E$ antenna. $d_1 + d_2 \leq \max\{N_T - N_E, 0\}$ then follows from the converse of Corollary 1.

## VI. DISCUSSION

### A. Methods of Proving Strong Secrecy

As we have mentioned in Section I, it is common practice to prove weak secrecy results and use privacy amplification [25] to arrive at strong secrecy with identical rates. Since this work deviates from that approach, and derives the strong secrecy rate directly, it is instructive to discuss the reason behind it.

Given a weak secrecy coding scheme that spans over $n$ channel uses, reference [25] considers an equivalent channel for which every $n$ inputs to the weak secrecy scheme is viewed as a single input to the equivalent channel. It then designs a strong secrecy scheme for this equivalent channel. If this approach were followed here, then the arbitrarily varying channel would be encapsulated inside a universal weak secrecy coding scheme to form the equivalent channel. However, this does not change the fact that the equivalent channel still has an arbitrarily varying joint distribution for its inputs and outputs. It can be shown that the resulting coding scheme does satisfy the strong secrecy constraint in (9) for all possible eavesdropper channel state sequences. However, the convergence speed of the limit in (9) is not uniform over these sequences. Due to this subtlety, the approach of [25] cannot not be used to prove strong secrecy in this work.

### B. Comparison with Related Work

In this section, we provide a detailed discussion of the differences of this work from previous work that considered settings similar or related to ours. These are:

- Reference [23]: This reference considered the MISO wiretap channel, and claimed that, with a certain probability, a positive *weak* secrecy rate can be guaranteed by introducing artificial noise at the transmitter, even if the transmitter has no knowledge about the eavesdropper channel states. However, as pointed out in [24], a *universal* coding scheme is necessary for this claim to be correct. In this work, we proved the existence of such a universal coding scheme.

- References [33] and [27]: The possibility of using information spectrum [34] to prove coding theorems for wiretap channels was first suggested in [33]. Recently [27] provided a more detailed proof based on this. The proof of [27] can be easily extended to the case of having a *finite* number of eavesdroppers, i.e., the compound setting. However, in this work, we have to consider infinitely possibilities for eavesdropper channel state sequences. This makes the proof in this work differs from [27] in the following aspects:

  1) To prove the results in [27], it is sufficient to prove that the left hand side of (72) converges to zero as $n$ goes to $\infty$. Here, we must prove it converges exponentially fast to zero with respect to $n$, as otherwise, the existence of universal secure coding scheme does *not* follow.

  2) In order to use the approximation argument from [35], we need a uniform upper bound on the average power of each codeword in the codebook. This was used in obtaining (88) from (87). To obtain such a bound, we have to sample from a truncated $n$-letter Gaussian distribution, as shown in (46) and (48), which complicates the analysis of the information density. In contrast, in the setting of [27], it is sufficient to sample from a single letter Gaussian distribution.

  The model considered by [27] is also inherently different from the model considered in this work in the sense that the sequences of eavesdropper channel states in [27] must have a well-defined $n$ letter distribution. We do not have this restriction in this work.

- Reference [24]: Recently, [24] considered a general arbitrarily varying *discrete* memoryless wiretap channel and proved the existence of a universal coding scheme for this channel for *weak* secrecy. This work differs from [24] in that we consider a Gaussian model, whose inputs and outputs are *continuous* and we prove *strong* secrecy. We stress that for the arbitrarily varying channel setting [38], the analysis for continuous alphabets do not follow from its discrete counterpart with finite alphabets [24, Lemma 3, Lemma 4]. Thus, care must be exercised in establishing the results from scratch as evidenced by Section IV.

## VII. Conclusion

In this work, we have considered secure communication in the presence of eavesdroppers whose channels are unknown to the legitimate parties and can be arbitrarily varying. We have shown that multiple antennas used in conjunction with the universal coding scheme presented in this paper can guarantee positive secrecy rates irrespective of the channels of the (possibly infinite numbers of) eavesdroppers.

We have derived achievable secrecy rates for the MIMO wiretap channel, and achievable secrecy rate regions for the MIMO MAC wiretap channel and the MIMO broadcast wiretap channel where each transmitter(s) and the intended receiver(s) have the same number of antennas. We have also derived the secure degrees of freedom, and the secure degrees of freedom regions for these channels by matching the converse to the achievable rates in high SNR. These results are guaranteed in the most stringent setting, that is in the sense of strong secrecy, which coincides with the results derived when the notion of weak secrecy is used and the eavesdropper's channel gain is assumed to be static.

As future work, it is of interest to consider MIMO MAC and MIMO BC wiretap channels with asymmetric number of antennas, for which a time sharing scheme is unlikely to be optimal. In this work, we considered the case where conditioned on a given sequence of channel states, the eavesdropper channel is memoryless. The corresponding channel model with memory deserves further investigation.

## Appendix A
## Proof of Lemma 2

The second term in (63) can be bounded as:

$$\int |f_{\gamma,\tilde{\mathbf{Y}}_G^n}(z^n) - f_{\gamma,\tilde{\mathbf{Y}}_T^n}(z^n)(z^n)|dz^n \tag{214}$$

$$\leq \int |\int f_{\gamma,\tilde{\mathbf{Y}}_G^n|\tilde{\mathbf{X}}_G^n}(z^n|x^n) f_{\tilde{\mathbf{X}}_G^n}(x^n) dx^n - \int f_{\gamma,\tilde{\mathbf{Y}}_T^n|\tilde{\mathbf{X}}_T^n}(z^n|x^n) f_{\mathbf{X}_T^n}(x^n) dx^n |dz^n \tag{215}$$

$$= \int |\int f_{\gamma,\tilde{\mathbf{Y}}_G^n|\tilde{\mathbf{X}}_G^n}(z^n|x^n) f_{\tilde{\mathbf{X}}_G^n}(x^n) dx^n - \int f_{\gamma,\tilde{\mathbf{Y}}_G^n|\tilde{\mathbf{X}}_G^n}(z^n|x^n) f_{\mathbf{X}_T^n}(x^n) dx^n |dz^n \tag{216}$$

$$\leq \int \int f_{\gamma,\tilde{\mathbf{Y}}_G^n|\tilde{\mathbf{X}}_G^n}(z^n|x^n) |f_{\tilde{\mathbf{X}}_G^n}(x^n) - f_{\mathbf{X}_T^n}(x^n)|dx^n dz^n \tag{217}$$

$$= \int \int f_{\gamma,\tilde{\mathbf{Y}}_G^n-\tilde{\mathbf{H}}^n\tilde{\mathbf{X}}_G^n}\left(z^n - \tilde{\mathbf{H}}^n x^n\right) |f_{\tilde{\mathbf{X}}_G^n}(x^n) - f_{\mathbf{X}_T^n}(x^n)|dx^n dz^n \tag{218}$$

Let $u^n = z^n - \tilde{\mathbf{H}}^n x^n$. Then (218) equals:[6]

$$\int \int f_{\gamma, \tilde{\mathbf{Y}}_G^n - \tilde{\mathbf{H}}^n \tilde{\mathbf{X}}_G^n}(u^n) |f_{\tilde{\mathbf{X}}_G^n}(x^n) - f_{\mathbf{X}_T^n}(x^n)| dx^n du^n \tag{219}$$

$$= \int f_{\gamma, \tilde{\mathbf{Y}}_G^n - \tilde{\mathbf{H}}^n \tilde{\mathbf{X}}_G^n}(u^n) \int |f_{\tilde{\mathbf{X}}_G^n}(x^n) - f_{\mathbf{X}_T^n}(x^n)| dx^n du^n \tag{220}$$

For the innermost integral in (220), we can write:

$$\int |f_{\tilde{\mathbf{X}}_G^n}(x^n) - f_{\mathbf{X}_T^n}(x^n)| dx^n \tag{221}$$

$$= \int_{\frac{1}{n}\|x^n\|^2 > P} |f_{\tilde{\mathbf{X}}_G^n}(x^n) - f_{\mathbf{X}_T^n}(x^n)| dx^n + \int_{\frac{1}{n}\|x^n\|^2 < P} |f_{\tilde{\mathbf{X}}_G^n}(x^n) - f_{\mathbf{X}_T^n}(x^n)| dx^n \tag{222}$$

$$\leq \int_{\frac{1}{n}\|x^n\|^2 > P} f_{\tilde{\mathbf{X}}_G^n}(x^n) dx^n + \int_{\frac{1}{n}\|x^n\|^2 > P} f_{\mathbf{X}_T^n}(x^n) dx^n$$

$$+ \int_{\frac{1}{n}\|x^n\|^2 < P} |f_{\tilde{\mathbf{X}}_G^n}(x^n) - f_{\mathbf{X}_T^n}(x^n)| dx^n \tag{223}$$

$$\leq (1 - \mu_{n,\varepsilon_P}) + \int_{\frac{1}{n}\|x\|^2 < P} |f_{\tilde{\mathbf{X}}_G^n}(x^n) - \mu_{n,\varepsilon_P}^{-1} f_{\tilde{\mathbf{X}}_G^n}(x^n)| dx^n \tag{224}$$

$$\leq (1 - \mu_{n,\varepsilon_P}) + \mu_{n,\varepsilon_P}^{-1} - 1 \tag{225}$$

$$= \mu_{n,\varepsilon_P}^{-1} - \mu_{n,\varepsilon_P} \tag{226}$$

From (49), we can choose sufficiently large $n$, such that $\mu_{n,\varepsilon_P} > 1/2$. For such $n$, we have

$$\mu_{n,\varepsilon_P}^{-1} - \mu_{n,\varepsilon_P} \tag{227}$$

$$= \mu_{n,\varepsilon_P}^{-1} \left(1 - \mu_{n,\varepsilon_P}^2\right) \tag{228}$$

$$\leq 2 \left(1 - \mu_{n,\varepsilon_P}^2\right) \tag{229}$$

$$= 2 \left(1 - \mu_{n,\varepsilon_P}\right)\left(1 + \mu_{n,\varepsilon_P}\right) \tag{230}$$

$$\leq 4 \left(1 - \mu_{n,\varepsilon_P}\right) \tag{231}$$

Therefore (220) is upper bounded by $4e^{-n\alpha(\varepsilon_P)}$. This concludes the proof of the lemma.

## APPENDIX B

### PROOF OF (66)

Recall that $\mathrm{E}_B[A]$ denotes the expectation of $A$ averaged over $B$. Define $L = 2^{n(I(\tilde{\mathbf{X}}; \tilde{\mathbf{Y}}) + \delta_n)}$. Recall that $x_{i,j}^n$ denotes the codeword in the codebook $\mathcal{C}$ that is labeled with $(i, j)$. Since $(i, j)$

---

[6] It is easy to verify that the absolute value of the Jacobian determinant is 1 in this case.

are uniformly distributed, we have:

$$E_{\mathcal{C}}\left[\sum_w p_W(w) \int |f_{\gamma,\tilde{\mathbf{Y}}_G^n}(z^n) - f_{\gamma,\tilde{\mathbf{Y}}_{\mathcal{C}}^n|W}(z^n|w)|dz^n\right] \qquad (232)$$

$$=E_{\mathcal{C}}\left[\sum_w p_W(w) \int |f_{\gamma,\tilde{\mathbf{Y}}_G^n}(z^n) - \frac{1}{L}\sum_{j=1}^L f_{\gamma,\tilde{\mathbf{Y}}_{\mathcal{C}}^n|\tilde{\mathbf{X}}_{\mathcal{C}}^n}(z^n|x_{w,j}^n)|dz^n\right] \qquad (233)$$

$$=\sum_w p_W(w) \int \prod_{j=1}^L f_{\tilde{\mathbf{X}}_T^n}(x_{w,j}^n) \int |f_{\gamma,\tilde{\mathbf{Y}}_G^n}(z^n) - \frac{1}{L}\sum_{j=1}^L f_{\gamma,\tilde{\mathbf{Y}}_{\mathcal{C}}^n|\tilde{\mathbf{X}}_{\mathcal{C}}^n}(z^n|x_{w,j}^n)|dz^n dx_{w,j,j=1...L}^n \qquad (234)$$

$$=\sum_w p_W(w) \int \prod_{j=1}^L f_{\tilde{\mathbf{X}}_T^n}(x_{1,j}^n) \int |f_{\gamma,\tilde{\mathbf{Y}}_G^n}(z^n) - \frac{1}{L}\sum_{j=1}^L f_{\gamma,\tilde{\mathbf{Y}}_{\mathcal{C}}^n|\tilde{\mathbf{X}}_{\mathcal{C}}^n}(z^n|x_{1,j}^n)|dz^n dx_{1,j,j=1...L}^n \qquad (235)$$

$$=\int \prod_{j=1}^L f_{\tilde{\mathbf{X}}_T^n}(x_{1,j}^n) \int |f_{\gamma,\tilde{\mathbf{Y}}_G^n}(z^n) - \frac{1}{L}\sum_{j=1}^L f_{\gamma,\tilde{\mathbf{Y}}_{\mathcal{C}}^n|\tilde{\mathbf{X}}_{\mathcal{C}}^n}(z^n|x_{1,j}^n)|dz^n dx_{1,j,j=1...L}^n \qquad (236)$$

$$=E_{\mathcal{C}}\left[\int |f_{\gamma,\tilde{\mathbf{Y}}_G^n}(z^n) - \frac{1}{L}\sum_{j=1}^L f_{\gamma,\tilde{\mathbf{Y}}_{\mathcal{C}}^n|\tilde{\mathbf{X}}_{\mathcal{C}}^n}(z^n|x_{1,j}^n)|dz^n\right] \qquad (237)$$

$$=E_{\mathcal{C}}\left[\int |f_{\gamma,\tilde{\mathbf{Y}}_G^n}(z^n) - f_{\gamma,\tilde{\mathbf{Y}}_{\mathcal{C}}^n|W}(z^n|1)|dz^n\right] \qquad (238)$$

which is (66).

## APPENDIX C
## PROOF OF LEMMA 4

Let $M$ be the number of codewords in each bin in the codebook:

$$M = 2^{n(I(\tilde{\mathbf{X}}_G;\tilde{\mathbf{Y}}_G)+\delta_n)} \qquad (239)$$

As in [34, Proof of Theorem 4], we begin by defining random variables $\tilde{\mathbf{X}}^n$, $\tilde{\mathbf{X}}_1^n,...,\tilde{\mathbf{X}}_M^n$ and $\tilde{\mathbf{Y}}^n$ such that

1) The distribution of $\tilde{\mathbf{X}}^n$ is also given by $Q_{\tilde{\mathbf{X}}^n}$.
2) The distribution of $\tilde{\mathbf{Y}}^n$ conditioned on $\tilde{\mathbf{X}}^n$ is determined by the eavesdropper channel.
3) $\tilde{\mathbf{X}}_1^n,...,\tilde{\mathbf{X}}_M^n$ are i.i.d. and the distribution of $\tilde{\mathbf{X}}_j^n, j=1,...,M$ is given by $Q_{\tilde{\mathbf{X}}^n}$. $\tilde{\mathbf{X}}_1^n,...,\tilde{\mathbf{X}}_M^n$ are independent from $\tilde{\mathbf{Y}}^n$.

Then, we have [34, Proof of Theorem 4, (4.2)]:

$$\mathbb{E}_{\mathcal{C}}\left[\Pr\left[\log\frac{f_{\gamma,\tilde{\mathbf{Y}}_{\mathcal{C}}^{n}|W}\left(\tilde{\mathbf{Y}}_{\mathcal{C}}^{n}|1\right)}{f_{\gamma,\tilde{\mathbf{Y}}_{G}^{n}}\left(\tilde{\mathbf{Y}}_{\mathcal{C}}^{n}\right)}>\mu_{n}\right]\right] \tag{240}$$

$$\leq \Pr\left[\frac{1}{M}2^{i_{\gamma,\tilde{\mathbf{X}}_{G}^{n}\tilde{\mathbf{Y}}_{G}^{n}}(\tilde{\mathbf{X}}^{n},\tilde{\mathbf{Y}}^{n})}>\tau_{n}\right]+\Pr\left[\frac{1}{M}\sum_{j=1}^{M}2^{i_{\gamma,\tilde{\mathbf{X}}_{G}^{n}\tilde{\mathbf{Y}}_{G}^{n}}(\tilde{\mathbf{X}}_{j}^{n},\tilde{\mathbf{Y}}^{n})}>c_{2,n}+\tau_{n}\right] \tag{241}$$

where $c_{2,n}$ and $\tau_n > 0$ satisfy

$$2\tau_n + c_{2,n} = 2^{\mu_n} \tag{242}$$

The values of $c_{2,n}$ and $\tau_n$ will be specified later.

Let $1\{a > b\}$ denote the indicator function that equals $1$ if $a > b$, $0$ otherwise. For the first term in (241), we can write:

$$\Pr\left[\frac{1}{M}2^{i_{\gamma,\tilde{\mathbf{X}}_{G}^{n}\tilde{\mathbf{Y}}_{G}^{n}}(\tilde{\mathbf{X}}^{n},\tilde{\mathbf{Y}}^{n})}>\tau_{n}\right] \tag{243}$$

$$=\int f_{\tilde{\mathbf{X}}^{n}}(x^{n})f_{\gamma,\tilde{\mathbf{Y}}^{n}|\tilde{\mathbf{X}}^{n}}(y^{n}|x^{n})\,1\left\{\frac{1}{M}2^{i_{\gamma,\tilde{\mathbf{X}}_{G}^{n}\tilde{\mathbf{Y}}_{G}^{n}}(x^{n},y^{n})}>\tau_{n}\right\}dx^{n}dy^{n} \tag{244}$$

$$\leq \mu_{n,\varepsilon_P}^{-1}\int f_{\tilde{\mathbf{X}}_G^{n}}(x^{n})f_{\gamma,\tilde{\mathbf{Y}}_G^{n}|\tilde{\mathbf{X}}_G^{n}}(y^{n}|x^{n})\,1\left\{\frac{1}{M}2^{i_{\gamma,\tilde{\mathbf{X}}_{G}^{n}\tilde{\mathbf{Y}}_{G}^{n}}(x^{n},y^{n})}>\tau_{n}\right\}dx^{n}dy^{n} \tag{245}$$

$$\leq \mu_{n,\varepsilon_P}^{-1}\Pr\left[\frac{1}{M}2^{i_{\gamma,\tilde{\mathbf{X}}_{G}^{n}\tilde{\mathbf{Y}}_{G}^{n}}(\tilde{\mathbf{X}}_G^{n},\tilde{\mathbf{Y}}_G^{n})}>\tau_{n}\right] \tag{246}$$

$$=\mu_{n,\varepsilon_P}^{-1}\Pr\left[\frac{1}{n}i_{\gamma,\tilde{\mathbf{X}}_{G}^{n}\tilde{\mathbf{Y}}_{G}^{n}}\left(\tilde{\mathbf{X}}_G^{n},\tilde{\mathbf{Y}}_G^{n}\right)>I\left(\tilde{\mathbf{X}}_G;\tilde{\mathbf{Y}}_G\right)+\delta_n+\frac{1}{n}\log_2\tau_n\right] \tag{247}$$

For the second term in (241), we follow [34, (4.4),(4.5)] and define the following random variables conditioned on $\tilde{\mathbf{Y}}^n = y^n$:

$$V_{n,j}(y^n) = 2^{i_{\gamma,\tilde{\mathbf{X}}_{G}^{n}\tilde{\mathbf{Y}}_{G}^{n}}(\tilde{\mathbf{X}}_j^n,y^n)} \tag{248}$$

$$Z_{n,j}(y^n) = V_{n,j}(y^n)\,1\{V_{n,j}(y^n) \leq M\} \tag{249}$$

$$U_M(y^n) = \frac{1}{M}\sum_{j=1}^{M}V_{n,j}(y^n) \tag{250}$$

$$T_M(y^n) = \frac{1}{M}\sum_{j=1}^{M}Z_{n,j}(y^n) \tag{251}$$

With these notations, as in [34, (4.6)], we can write:

$$\Pr\left[\frac{1}{M}\sum_{j=1}^{M}2^{i_{\gamma,\tilde{\mathbf{X}}_{G}^{n}\tilde{\mathbf{Y}}_{G}^{n}}(\tilde{\mathbf{X}}^n,\tilde{\mathbf{Y}}^n)}>c_{2,n}+\tau_n\Big|\tilde{\mathbf{Y}}^n=y^n\right] \tag{252}$$

$$= \Pr\left[U_M(y^n) > c_{2,n} + \tau_n\right] \tag{253}$$

$$\leq \Pr\left(T_M(y^n) \neq U_M(y^n)\right) + \Pr\left(T_M(y^n) > c_{2,n} + \tau_n\right) \tag{254}$$

For the first term in (254), we can write [34]:

$$\Pr\left(T_M(y^n) \neq U_M(y^n)\right) \tag{255}$$

$$\leq \sum_{j=1}^{M} \Pr\left[Z_{n,j}(y^n) \neq V_{n,j}(y^n)\right] \tag{256}$$

$$= M \Pr\left[V_{n,1}(y^n) > M\right] \tag{257}$$

(257) can be upper bounded as:

$$M\left[\Pr\left[V_{n,1}\left(\tilde{\mathbf{Y}}^n\right) > M\right]\right] \tag{258}$$

$$= M\left[\Pr\left[2^{i_{\gamma,\tilde{\mathbf{X}}_G^n \tilde{\mathbf{Y}}_G^n}(\tilde{\mathbf{X}}^n,\tilde{\mathbf{Y}}^n)} > M\right]\right] \tag{259}$$

$$= M \int f_{\tilde{\mathbf{X}}^n}(x^n) f_{\gamma,\tilde{\mathbf{Y}}^n|\tilde{\mathbf{X}}^n}(z^n|x^n) \mathbf{1}\left\{2^{i_{\gamma,\tilde{\mathbf{X}}_G^n \tilde{\mathbf{Y}}_G^n}(x^n,z^n)} > M\right\} dx^n dz^n \tag{260}$$

$$\leq \mu_{n,\varepsilon_P}^{-1} M \int f_{\tilde{\mathbf{X}}_G^n}(x^n) f_{\gamma,\tilde{\mathbf{Y}}_G^n|\tilde{\mathbf{X}}_G^n}(z^n|x^n) \mathbf{1}\left\{2^{i_{\gamma,\tilde{\mathbf{X}}_G^n \tilde{\mathbf{Y}}_G^n}(x^n,z^n)} > M\right\} dx^n dz^n \tag{261}$$

As shown in [34, Proof of Theorem 4], (261) is upper bounded by:

$$\mu_{n,\varepsilon_P}^{-1} \Pr\left[\frac{1}{n} i_{\gamma,\tilde{\mathbf{X}}_G^n \tilde{\mathbf{Y}}_G^n}\left(\tilde{\mathbf{X}}_G^n, \tilde{\mathbf{Y}}_G^n\right) > I\left(\tilde{\mathbf{X}}_G; \tilde{\mathbf{Y}}_G\right) + \delta_n\right] \tag{262}$$

This means

$$\Pr\left(T_M\left(\tilde{\mathbf{Y}}^n\right) \neq U_M\left(\tilde{\mathbf{Y}}^n\right)\right) \tag{263}$$

$$\leq \mu_{n,\varepsilon_P}^{-1} \Pr\left[\frac{1}{n} i_{\gamma,\tilde{\mathbf{X}}_G^n \tilde{\mathbf{Y}}_G^n}\left(\tilde{\mathbf{X}}_G^n, \tilde{\mathbf{Y}}_G^n\right) > I\left(\tilde{\mathbf{X}}_G; \tilde{\mathbf{Y}}_G\right) + \delta_n\right] \tag{264}$$

For the second term in (254), we can write [34, (4.7)]:

$$\mathrm{E}\left[T_M(y^n)\right] = \mathrm{E}\left[Z_{n,1}(y^n)\right] = \int f_{\tilde{\mathbf{X}}^n}(x^n) 2^{i_{\gamma,\tilde{\mathbf{X}}_G^n \tilde{\mathbf{Y}}_G^n}(x^n,y^n)} \mathbf{1}\left\{2^{i_{\gamma,\tilde{\mathbf{X}}_G^n \tilde{\mathbf{Y}}_G^n}(x^n,y^n)} \leq M\right\} dx^n \tag{265}$$

$$\leq \mu_{n,\varepsilon_P}^{-1} \int f_{\tilde{\mathbf{X}}_G^n}(x^n) 2^{i_{\gamma,\tilde{\mathbf{X}}_G^n \tilde{\mathbf{Y}}_G^n}(x^n,y^n)} \mathbf{1}\left\{2^{i_{\gamma,\tilde{\mathbf{X}}_G^n \tilde{\mathbf{Y}}_G^n}(x^n,y^n)} \leq M\right\} dx^n \tag{266}$$

$$\leq \mu_{n,\varepsilon_P}^{-1} \tag{267}$$

We choose $c_{2,n}$ as

$$c_{2,n} = \mu_{n,\varepsilon_p}^{-1} \tag{268}$$

Then, following [34, (4.8)], we have

$$\Pr\left(T_M\left(y^n\right) > c_{2,n} + \tau_n\right) \leq \Pr\left(T_M\left(y^n\right) - \mathrm{E}\left[T_M\left(y^n\right)\right] > \tau_n\right) \quad (269)$$

$$\leq \frac{1}{\tau_n^2}\mathrm{var}\left(T_M\left(y^n\right)\right) \quad (270)$$

$$\leq \frac{1}{\tau_n^2}\mathrm{E}\left[\frac{1}{M}Z_{n,1}^2\left(y^n\right)\right] \quad (271)$$

The expectation in (271) can be upper bounded by:

$$\mathrm{E}\left[\frac{1}{M}Z_{n,1}^2\left(y^n\right)\right] = \frac{1}{M}\int f_{\tilde{\mathbf{X}}^n}(x^n) f_{\gamma,\tilde{\mathbf{Y}}^n|\tilde{\mathbf{X}}^n}(y^n|x^n) Z_{n,1}^2(y^n)\,dx^n dy^n \quad (272)$$

$$= \frac{1}{M}\int f_{\tilde{\mathbf{X}}^n}(x^n) f_{\gamma,\tilde{\mathbf{Y}}^n|\tilde{\mathbf{X}}^n}(y^n|x^n)(V_{n,1}(y^n)\mathbf{1}\{V_{n,1}(y^n) \leq M\})^2 dx^n dy^n \quad (273)$$

$$= \frac{1}{M}\int f_{\tilde{\mathbf{X}}^n}(x^n) f_{\gamma,\tilde{\mathbf{Y}}^n|\tilde{\mathbf{X}}^n}(y^n|x^n)(2^{i_{\gamma,\tilde{\mathbf{X}}_G^n\tilde{\mathbf{Y}}_G^n}(x^n,y^n)}\mathbf{1}\{2^{i_{\gamma,\tilde{\mathbf{X}}_G^n\tilde{\mathbf{Y}}_G^n}(x^n,y^n)} \leq M\})^2 dx^n dy^n \quad (274)$$

$$\leq \mu_{n,\varepsilon_P}^{-1}\frac{1}{M}\int f_{\tilde{\mathbf{X}}_G^n}(x^n) f_{\gamma,\tilde{\mathbf{Y}}_G^n|\tilde{\mathbf{X}}_G^n}(y^n|x^n)(2^{i_{\gamma,\tilde{\mathbf{X}}_G^n\tilde{\mathbf{Y}}_G^n}(x^n,y^n)}\mathbf{1}\{2^{i_{\gamma,\tilde{\mathbf{X}}_G^n\tilde{\mathbf{Y}}_G^n}(x^n,y^n)} \leq M\})^2 dx^n dy^n \quad (275)$$

As illustrated in [34, Proof of Theorem 4,], equation (275) is upper bounded by:

$$\mu_{n,\varepsilon_P}^{-1}\{2^{-n\frac{\delta_n}{2}} + \Pr\left[\frac{1}{n}i_{\gamma,\tilde{\mathbf{X}}_G^n\tilde{\mathbf{Y}}_G^n}\left(\tilde{\mathbf{X}}_G^n,\tilde{\mathbf{Y}}_G^n\right) > I\left(\tilde{\mathbf{X}}_G;\tilde{\mathbf{Y}}_G\right) + \frac{\delta_n}{2}\right]\} \quad (276)$$

This means

$$\Pr\left(T_M\left(\tilde{\mathbf{Y}}^n\right) > c_{2,n} + \tau_n\right) \quad (277)$$

$$\leq \frac{\mu_{n,\varepsilon_P}^{-1}}{\tau_n^2}\{2^{-n\frac{\delta_n}{2}} + \Pr\left[\frac{1}{n}i_{\gamma,\tilde{\mathbf{X}}_G^n\tilde{\mathbf{Y}}_G^n}\left(\tilde{\mathbf{X}}_G^n,\tilde{\mathbf{Y}}_G^n\right) > I\left(\tilde{\mathbf{X}}_G;\tilde{\mathbf{Y}}_G\right) + \frac{\delta_n}{2}\right]\} \quad (278)$$

Substituting (263)-(264) and (277)-(278) to (252)-(254), we observe:

$$\Pr\left[\frac{1}{M}\sum_{j=1}^M 2^{i_{\gamma,\tilde{\mathbf{X}}_G^n\tilde{\mathbf{Y}}_G^n}\left(\tilde{\mathbf{X}}^n,\tilde{\mathbf{Y}}^n\right)} > c_{2,n} + \tau_n\right] \quad (279)$$

$$\leq \mu_{n,\varepsilon_P}^{-1}\Pr\left[\frac{1}{n}i_{\gamma,\tilde{\mathbf{X}}_G^n\tilde{\mathbf{Y}}_G^n}\left(\tilde{\mathbf{X}}_G^n,\tilde{\mathbf{Y}}_G^n\right) > I\left(\tilde{\mathbf{X}}_G;\tilde{\mathbf{Y}}_G\right) + \delta_n\right] +$$

$$\frac{\mu_{n,\varepsilon_P}^{-1}}{\tau_n^2}\{2^{-n\frac{\delta_n}{2}} + \Pr\left[\frac{1}{n}i_{\gamma,\tilde{\mathbf{X}}_G^n\tilde{\mathbf{Y}}_G^n}\left(\tilde{\mathbf{X}}_G^n,\tilde{\mathbf{Y}}_G^n\right) > I\left(\tilde{\mathbf{X}}_G;\tilde{\mathbf{Y}}_G\right) + \frac{\delta_n}{2}\right]\} \quad (280)$$

Applying this result along with (243)-(247) to (240)-(241), we obtain Lemma 4.

# APPENDIX D

## PROOF OF LEMMA 5

Define $P'$ as

$$P' = \frac{P(1-\varepsilon_P)}{N_T} + 1 \tag{281}$$

We prove Lemma 5 when $\tilde{\mathbf{H}}^n$ is a sequence such that $\tilde{\mathbf{H}}(i)$ is given by (36). The case where $\tilde{\mathbf{H}}(i)$ is invariant with respect to $i$ is a special case, and does not require a separate proof.

We begin with:

$$\prod_{i=1}^n f_{\gamma,\tilde{\mathbf{Y}}|\tilde{\mathbf{X}}}\left(\tilde{\mathbf{Y}}_i|\tilde{\mathbf{X}}_i\right) = \frac{1}{(\pi)^{nN_E}} \exp\left(-\left\|\tilde{\mathbf{Y}}^n - \tilde{\mathbf{H}}^n\tilde{\mathbf{X}}^n\right\|^2\right) \tag{282}$$

where we used the notation in (32). Note that with the distribution we choose for $\tilde{\mathbf{X}}_G^n$ and the channel matrix sequence $\tilde{\mathbf{H}}^n$ given by (36), $\tilde{\mathbf{Y}}_G^n$ is a rotationally invariant complex Gaussian random vector with zero mean and covariance matrix $P'\mathbf{I}$:

$$f_{\gamma,\tilde{\mathbf{Y}}_G^n}\left(\tilde{\mathbf{Y}}^n\right) = \frac{1}{(\pi P')^{nN_E}} \exp\left(-\frac{\left\|\tilde{\mathbf{Y}}^n\right\|^2}{P'}\right) \tag{283}$$

Therefore

$$\frac{1}{n} i_{\gamma,\tilde{\mathbf{X}}_G^n\tilde{\mathbf{Y}}_G^n}\left(\tilde{\mathbf{X}}_G^n, \tilde{\mathbf{Y}}_G^n\right)$$
$$= N_E \log_2(P') + \left\{\frac{1}{n}\left(\frac{\left\|\tilde{\mathbf{Y}}_G^n\right\|^2}{P'}\right) - \frac{1}{n}\left(\left\|\tilde{\mathbf{Y}}_G^n - \tilde{\mathbf{H}}^n\tilde{\mathbf{X}}_G^n\right\|^2\right)\right\} \log_2 e \tag{284}$$

Define $\tilde{\mathbf{N}}^n$ as

$$\tilde{\mathbf{N}}^n = \tilde{\mathbf{Y}}_G^n - \tilde{\mathbf{H}}^n\tilde{\mathbf{X}}_G^n \tag{285}$$

For a fixed positive constant $\varepsilon_2$, we have:

$$\Pr(\frac{1}{n} i_{\gamma,\tilde{\mathbf{X}}_G^n\tilde{\mathbf{Y}}_G^n}(\tilde{\mathbf{X}}_G^n, \tilde{\mathbf{Y}}_G^n) > I(\tilde{\mathbf{X}}_G; \tilde{\mathbf{Y}}_G) + \delta_n) \tag{286}$$

$$= \Pr(N_E \log_2(P') + \{\frac{1}{n}(\frac{\left\|\tilde{\mathbf{Y}}_G^n\right\|^2}{P'}) - \frac{1}{n}\left\|\tilde{\mathbf{N}}^n\right\|^2\} \log_2 e > I(\tilde{\mathbf{X}}_G; \tilde{\mathbf{Y}}_G) + \delta_n) \tag{287}$$

$$= \Pr(N_E \log_2(P') + \frac{\log_2 e}{n}(\frac{\left\|\tilde{\mathbf{Y}}_G^n\right\|^2}{P'}) > I(\tilde{\mathbf{X}}_G; \tilde{\mathbf{Y}}_G) + \frac{\log_2 e}{n}\left\|\tilde{\mathbf{N}}^n\right\|^2 + \delta_n) \tag{288}$$

$$\leq \Pr(\frac{1}{n}\left\|\tilde{\mathbf{N}}^n\right\|^2 < N_E(1-\varepsilon_2)) + \Pr(\frac{1}{n}\left\|\tilde{\mathbf{N}}^n\right\|^2 \geq N_E(1-\varepsilon_2))$$

$$\Pr(N_E \log_2(P') + \frac{\log_2 e}{n}(\frac{\left\|\tilde{\mathbf{Y}}_G^n\right\|^2}{P'}) > I(\tilde{\mathbf{X}}_G; \tilde{\mathbf{Y}}_G) +$$

$$\frac{\log_2 e}{n}\left\|\tilde{\mathbf{N}}^n\right\|^2 + \delta_n \left|\frac{1}{n}\left\|\tilde{\mathbf{N}}^n\right\|^2 \geq N_E(1-\varepsilon_2)) \quad (289)$$

$$\leq \Pr(\frac{1}{n}\left\|\tilde{\mathbf{N}}^n\right\|^2 < N_E(1-\varepsilon_2))$$

$$+ \Pr(N_E \log_2(P') + \frac{\log_2 e}{n}(\frac{\left\|\tilde{\mathbf{Y}}_G^n\right\|^2}{P'}) > I(\tilde{\mathbf{X}}_G; \tilde{\mathbf{Y}}_G) + (\log_2 e)N_E(1-\varepsilon_2) + \delta_n$$

$$\left|\frac{1}{n}\left\|\tilde{\mathbf{N}}^n\right\|^2 \geq N_E(1-\varepsilon_2)) \quad (290)$$

$$= \Pr(\frac{1}{n}\left\|\tilde{\mathbf{N}}^n\right\|^2 < N_E(1-\varepsilon_2))$$

$$+ \Pr(N_E \log_2(P') + \frac{\log_2 e}{n}(\frac{\left\|\tilde{\mathbf{Y}}_G^n\right\|^2}{P'}) > I(\tilde{\mathbf{X}}_G; \tilde{\mathbf{Y}}_G) + (\log_2 e)N_E(1-\varepsilon_2) + \delta_n) \quad (291)$$

Note that $\tilde{\mathbf{N}}^n$ is a zero mean Gaussian random vector whose covariance matrix is $\mathbf{I}$. Hence, from [32, (B1)], there exists $\alpha(\varepsilon_2)$, such that

$$\Pr(\frac{1}{nN_E}\left\|\tilde{\mathbf{N}}^n\right\|^2 < 1-\varepsilon_2) < e^{-n\alpha(\varepsilon_2)} \quad (292)$$

In addition, note that

$$I(\tilde{\mathbf{X}}_G; \tilde{\mathbf{Y}}_G) - N_E \log_2(P') \quad (293)$$

$$= N_E \log_2(1 + \frac{P(1-\varepsilon_P)}{N_T}) - N_E \log_2(1 + \frac{P(1-\varepsilon_P)}{N_T}) = 0 \quad (294)$$

Hence, the second term in (291) can be written as:

$$\Pr(\frac{1}{n}\frac{\left\|\tilde{\mathbf{Y}}_G^n\right\|^2}{P'} > N_E(1-\varepsilon_2) + \frac{\delta_n}{\log_2 e}) = \Pr(\frac{1}{nN_E}\left\|\frac{\tilde{\mathbf{Y}}_G^n}{\sqrt{P'}}\right\|^2 > 1-\varepsilon_2 + \frac{\delta_n}{N_E \log_2 e}) \quad (295)$$

Each component of $\frac{\tilde{\mathbf{Y}}_G^n}{\sqrt{P'}}$ is a rotationally invariant zero mean complex Gaussian random variable with unit variance, regardless of the value of $\tilde{\mathbf{H}}^n$ and these components are independent.

Therefore, for $\varepsilon_3 > 0$, we have: if for all $n$, $1 - \varepsilon_2 + \frac{\delta_n}{N_E \log_2 e} \geq 1 + \varepsilon_3$, i.e., $\delta_n \geq N_E(\varepsilon_3 + \varepsilon_2)\log_2 e$, there must exist $\alpha(\varepsilon_3)$, such that [32, (B2)]:

$$\Pr\left(\frac{1}{nN_E}\left\|\frac{\tilde{\mathbf{Y}}_G^n}{\sqrt{P'}}\right\|^2 > 1 - \varepsilon_2 + \frac{\delta_n}{N_E \log_2 e}\right) < e^{-n\alpha(\varepsilon_3)} \quad (296)$$

We have obtained an exponential bound on both terms of (291). Finally, we let $\varepsilon_2 = \varepsilon_3$ and $\varepsilon = 2N_E \varepsilon_2 \log_2 e$. Hence $\delta_n \geq \varepsilon$, and we obtain Lemma 5.

## APPENDIX E

## PROOF OF LEMMA 6

Combining (242) and (268), we have:

$$\tau_n = \frac{e^{\mu_n \ln 2} - \mu_{n,\varepsilon_p}^{-1}}{2} \tag{297}$$

Since $\mu_n > 0$, we find that (297) is lower bounded by:

$$\frac{\mu_n \ln 2 + 1 - \mu_{n,\varepsilon_p}^{-1}}{2} \tag{298}$$

From (49), we can choose a sufficiently large $n$ such that $\mu_{n,\varepsilon_p} > 1/2$. This means $\mu_{n,\varepsilon_p}^{-1} < 2$. Since from (49), $1 - \mu_{n,\varepsilon_P} \leq e^{-n\alpha(\varepsilon_P)}$, we have

$$\mu_{n,\varepsilon_p}^{-1} - 1 < 2e^{-n\alpha(\varepsilon_P)} \tag{299}$$

Applying (299) to (298), we have

$$\tau_n \geq \frac{\mu_n \ln 2 - 2e^{-n\alpha(\varepsilon_P)}}{2} \tag{300}$$

The remainder of the proof entails finding an upper bound for (71), which will lead to an upper bound on (74) via Lemma 3. (71) can be bounded using Lemma 5 if its conditions are satisfied. This means that, for a given $\varepsilon > 0$, we require the following three conditions to be satisfied for all $n$.

$$\frac{\delta_n}{2} \geq \varepsilon \tag{301}$$

$$\delta_n + \frac{1}{n} \log_2 \tau_n \geq \varepsilon \tag{302}$$

$$\tau_n > 0 \tag{303}$$

Suppose all three conditions are fulfilled, then the three terms in (71) can be bounded as follows. The third term, as shown in Lemma 5 and using (297)-(298), can be bounded as

$$\frac{1}{\tau_n^2} \left( \Pr \left[ \frac{1}{n} i_{\gamma, \tilde{\mathbf{X}}_G^n \tilde{\mathbf{Y}}_G^n} \left( \tilde{\mathbf{X}}_G^n, \tilde{\mathbf{Y}}_G^n \right) > I\left( \tilde{\mathbf{X}}_G; \tilde{\mathbf{Y}}_G \right) + \frac{\delta_n}{2} \right] + 2^{-n\delta_n/2} \right) \tag{304}$$

$$\leq \frac{4}{(\mu_n \ln 2 + 1 - \mu_{n,\varepsilon_p}^{-1})^2} (e^{-n\alpha'(\varepsilon)} + 2^{-n\delta_n/2}) \tag{305}$$

On the other hand, if (301)-(303) hold, the first two terms in Lemma 4 are all bounded by $e^{-n\alpha'(\varepsilon)}$. Therefore, from Lemma 4, we find

$$\mathrm{E}_{\mathcal{C}}\left[\Pr\left[\log_2\frac{f_{\gamma,\tilde{\mathbf{Y}}_{\mathcal{C}}^n|W}\left(\tilde{\mathbf{Y}}_{\mathcal{C}}^n|1\right)}{f_{\gamma,\tilde{\mathbf{Y}}_G^n}\left(\tilde{\mathbf{Y}}_{\mathcal{C}}^n\right)} > \mu_n\right]\right] \tag{306}$$

$$\leq \mu_{n,\varepsilon_p}^{-1}\{2e^{-n\alpha'(\varepsilon)} + \frac{4}{(\mu_n\ln 2 + 1 - \mu_{n,\varepsilon_p}^{-1})^2}(e^{-n\alpha'(\varepsilon)} + 2^{-n\delta_n/2})\} \tag{307}$$

Then, from Lemma 3, we have

$$\mathrm{E}_{\mathcal{C}}\left[\int_{z^n}|f_{\gamma,\tilde{\mathbf{Y}}_G^n}(z^n) - f_{\gamma,\tilde{\mathbf{Y}}_{\mathcal{C}}^n|W}(z^n|1)|dz^n\right] \tag{308}$$

$$\leq \frac{2}{\log_2 e}\mu_n + 2\mu_{n,\varepsilon_p}^{-1}(2e^{-n\alpha'(\varepsilon)} + \frac{4}{(\mu_n\ln 2 + 1 - \mu_{n,\varepsilon_p}^{-1})^2}(e^{-n\alpha'(\varepsilon)} + 2^{-n\delta_n/2})) \tag{309}$$

From Lemma 2, we have

$$\int |f_{\gamma,\tilde{\mathbf{Y}}_G^n}(z^n) - f_{\gamma,\tilde{\mathbf{Y}}_T^n}(z^n)|dz^n < 4e^{-n\alpha(\varepsilon_P)} \tag{310}$$

Finally, from Lemma 1, we have

$$\mathrm{E}_{\mathcal{C}}[d_{\gamma,\mathcal{C}}] \leq \mathrm{E}_{\mathcal{C}}[2\sum_w p_W(w)\int_{z^n}|f_{\gamma,\tilde{\mathbf{Y}}_T^n}(z^n) - f_{\gamma,\tilde{\mathbf{Y}}_{\mathcal{C}}^n|W}(z^n|w)|dz^n] \tag{311}$$

$$\leq \mathrm{E}_{\mathcal{C}}[2\sum_w p_W(w)\int_{z^n}|f_{\gamma,\tilde{\mathbf{Y}}_G^n}(z^n) - f_{\gamma,\tilde{\mathbf{Y}}_{\mathcal{C}}^n|W}(z^n|w)|dz^n]$$

$$+ 2\int|f_{\gamma,\tilde{\mathbf{Y}}_G^n}(z^n) - f_{\gamma,\tilde{\mathbf{Y}}_T^n}(z^n)|dz^n \tag{312}$$

$$\leq \frac{4}{\log_2 e}\mu_n + 4\mu_{n,\varepsilon_p}^{-1}(2e^{-n\alpha'(\varepsilon)} + \frac{4}{(\mu_n\ln 2 + 1 - \mu_{n,\varepsilon_p}^{-1})^2}(e^{-n\alpha'(\varepsilon)} + e^{-n\delta_n \ln 2/2}))$$

$$+ 8e^{-n\alpha(\varepsilon_P)} \tag{313}$$

We then choose $\delta_n$ such that (301)-(303) hold. In order for the bound given by (313) to be small, we choose $\mu_n$ such that it decreases exponentially fast with respect to $n$. In order for the second term in (313) to decrease exponentially fast with respect to $n$, we can choose

$$\mu_n = e^{-n\min\{\frac{\alpha'(\varepsilon)}{4}, \frac{\delta_n \ln 2}{8}, \frac{\alpha(\varepsilon_P)}{2}\}} \tag{314}$$

The last term inside the minimum, $\alpha(\varepsilon_P)/2$, is for $\tau_n$ to remain positive, as required by (303). And as shown by (300), this means for sufficiently large $n$, whose lower bound only depends on $\varepsilon_P$, we have:

$$\tau_n \geq \frac{1}{4}e^{-n\frac{\alpha(\varepsilon_P)}{2}} \tag{315}$$

or equivalently:

$$-\frac{1}{n}\log_2 \tau_n \leq \frac{\alpha(\varepsilon_P)}{2}\log_2 e \tag{316}$$

Hence, we can choose $\delta_n$ such that $\delta_n \geq \max\{2\varepsilon, \varepsilon + \frac{\alpha(\varepsilon_P)}{2}\log_2 e\}$, then (301) and (302) hold.

Finally, for the above choices of $\mu_n$ and $\delta_n$, we note that both (313) and (309) decrease exponentially fast with respect to $n$. Hence, we have the two inequalities stated in Lemma 6.

## APPENDIX F
## PROOF OF LEMMA 7

As we had for Lemma 5, we prove Lemma 7 when $\tilde{\mathbf{H}}^n$ is a sequence such that $\tilde{\mathbf{H}}(i)$ is given by (36) and the eavesdropper channel is given by (41).

Let $f_\gamma$ be the conditional p.d.f. of the eavesdropper channel implied by (41) when the channel matrix sequence is $\tilde{\mathbf{H}}_\gamma^n$.

When $\frac{1}{n}\left\|\tilde{\mathbf{Y}}^n - \tilde{\mathbf{H}}_\gamma^n \tilde{\mathbf{X}}^n\right\|^2 < r^2$, for $\tilde{\mathbf{Y}}^n = z^n$ and $\tilde{\mathbf{X}}^n = x^n$, we have:

$$|\log f_\gamma(z^n|x^n) - \log f_{\gamma'}(z^n|x^n)| \tag{317}$$

$$= \left|\left\|z^n - \tilde{\mathbf{H}}_\gamma^n x^n\right\|^2 - \left\|z^n - \tilde{\mathbf{H}}_{\gamma'}^n x^n\right\|^2\right| \tag{318}$$

Recall that $\Delta_h^n = \tilde{\mathbf{H}}_\gamma^n - \tilde{\mathbf{H}}_{\gamma'}^n$ and $\mathbf{A}_i$ denotes the $i$th row of matrix $\mathbf{A}$. For a matrix $\mathbf{A}$ that has $N$ rows, let $[\mathbf{A}_1, ..., \mathbf{A}_N]$ denote a row vector formed by concatenating all rows of $\mathbf{A}$. Let $\langle x, y \rangle$ denote the inner product operation for the complex vector space. Note that $z^n$ is a $N_E \times n$ matrix here. Hence (318) can be upper bounded by:

$$2|\mathrm{Re}\sum_{i=1}^{N_E}\left\langle z_i^n - \tilde{\mathbf{H}}_{\gamma,i}^n x^n, \Delta_{h,i}^n x^n\right\rangle| + \sum_{i=1}^{N_E}\left\|\Delta_{h,i}^n x^n\right\|^2 \tag{319}$$

$$\leq 2|\sum_{i=1}^{N_E}\left\langle z_i^n - \tilde{\mathbf{H}}_{\gamma,i}^n x^n, \Delta_{h,i}^n x^n\right\rangle| + \sum_{i=1}^{N_E}\left\|\Delta_{h,i}^n x^n\right\|^2 \tag{320}$$

$$= 2|\left\langle [(z_1^n - \tilde{\mathbf{H}}_{\gamma,1}^n x^n), ..., (z_{N_E}^n - \tilde{\mathbf{H}}_{\gamma,N_E}^n x^n)], [\Delta_{h,1}^n x^n, ..., \Delta_{h,N_E}^n x^n]\right\rangle|$$
$$+ \sum_{i=1}^{N_E}\left\|\Delta_{h,i}^n x^n\right\|^2 \tag{321}$$

$$\leq 2\sqrt{\left\|z^n - \tilde{\mathbf{H}}_\gamma^n x^n\right\|^2 \|\Delta_x^n\|^2} + \|\Delta_x^n\|^2 \tag{322}$$

where in (322) we use the Cauchy-Schwartz inequality. From (89), we observe that (322) is upper bounded by:

$$ng(r, r') = nr'(2r + r') \tag{323}$$

From Lemma 1, (62)-(63) and Lemma 2, we have:

$$d_{\gamma,\mathcal{C}} \leq 2 \sum_w p_W(w) \int |f_{\gamma,\tilde{\mathbf{Y}}_T^n}(z^n) - f_{\gamma,\tilde{\mathbf{Y}}_\mathcal{C}^n|W}(z^n|w)| dz^n \tag{324}$$

$$\leq 2 \sum_w p_W(w) \int |f_{\gamma,\tilde{\mathbf{Y}}_G^n}(z^n) - f_{\gamma,\tilde{\mathbf{Y}}_\mathcal{C}^n|W}(z^n|w)| dz^n + 8e^{-n\alpha(\varepsilon_P)} \tag{325}$$

Hence we have obtained (95) in Lemma 7.

Recall that $\tilde{\mathbf{Y}}_G^n$ is the signal received by the eavesdropper if $\tilde{\mathbf{X}}^n = \tilde{\mathbf{X}}_G^n$. Since the input distribution we choose for $\tilde{\mathbf{X}}^n$ is zero mean rotationally invariant Gaussian with covariance matrix $(P(1 - \varepsilon_P))\mathbf{I}_{N_T \times N_T}$, and $\tilde{\mathbf{H}}_\gamma(i)$ always has the form given by (36) for all $\gamma$, we have

$$f_{\gamma,\tilde{\mathbf{Y}}_G^n}(z^n) = f_{\gamma',\tilde{\mathbf{Y}}_G^n}(z^n) \tag{326}$$

Therefore, the first term in (325) can be written as:

$$2 \sum_w p_W(w) \int |f_{\gamma',\tilde{\mathbf{Y}}_G^n}(z^n) - f_{\gamma,\tilde{\mathbf{Y}}_\mathcal{C}^n|W}(z^n|w)| dz^n \tag{327}$$

$$\leq 2 \sum_w p_W(w) \int |f_{\gamma,\tilde{\mathbf{Y}}_\mathcal{C}^n|W}(z^n|w) - f_{\gamma',\tilde{\mathbf{Y}}_\mathcal{C}^n|W}(z^n|w)| dz^n$$

$$+ 2 \sum_w p_W(w) \int |f_{\gamma',\tilde{\mathbf{Y}}_G^n}(z^n) - f_{\gamma',\tilde{\mathbf{Y}}_\mathcal{C}^n|W}(z^n|w)| dz^n \tag{328}$$

Recall that we label each codebook with $(i, j)$. In the encoder, we let $W = i$, the distribution over $j$ be $p_j$, which is uniform. We denote the codeword with label $(w, j)$ by $x_{w,j}^n$. Then, each term inside the sum over $w$ in the first term of (328) can be upper bounded as:

$$\int |f_{\gamma,\tilde{\mathbf{Y}}_\mathcal{C}^n|W}(z^n|w) - f_{\gamma',\tilde{\mathbf{Y}}_\mathcal{C}^n|W}(z^n|w)| dz^n \tag{329}$$

$$\leq \sum_j p_j \int |f_{\gamma,\tilde{\mathbf{Y}}^n|\tilde{\mathbf{X}}^n}(z^n|x_{w,j}^n) - f_{\gamma',\tilde{\mathbf{Y}}^n|\tilde{\mathbf{X}}^n}(z^n|x_{w,j}^n)| dz^n \tag{330}$$

The term inside the sum over $j$ can be upper bounded by:

$$\int |f_{\gamma,\tilde{\mathbf{Y}}^n|\tilde{\mathbf{X}}^n}(z^n|x_{w,j}^n) - f_{\gamma',\tilde{\mathbf{Y}}^n|\tilde{\mathbf{X}}^n}(z^n|x_{w,j}^n)| dz^n \tag{331}$$

$$= \int_{\frac{1}{n}\|z^n - \tilde{\mathbf{H}}_\gamma^n x_{w,j}^n\|^2 > r^2} |f_{\gamma,\tilde{\mathbf{Y}}^n|\tilde{\mathbf{X}}^n}(z^n|x_{w,j}^n) - f_{\gamma',\tilde{\mathbf{Y}}^n|\tilde{\mathbf{X}}^n}(z^n|x_{w,j}^n)| dz^n$$

$$+ \int_{\frac{1}{n}\|z^n - \tilde{\mathbf{H}}_\gamma^n x_{w,j}^n\|^2 < r^2} |f_{\gamma,\tilde{\mathbf{Y}}^n|\tilde{\mathbf{X}}^n}(z^n|x_{w,j}^n) - f_{\gamma',\tilde{\mathbf{Y}}^n|\tilde{\mathbf{X}}^n}(z^n|x_{w,j}^n)| dz^n \tag{332}$$

$$\leq \int_{\frac{1}{n}\|z^n - \tilde{\mathbf{H}}_\gamma^n x_{w,j}^n\|^2 > r^2} f_{\gamma,\tilde{\mathbf{Y}}^n|\tilde{\mathbf{X}}^n}(z^n|x_{w,j}^n) dz^n + \int_{\frac{1}{n}\|z^n - \tilde{\mathbf{H}}_\gamma^n x_{w,j}^n\|^2 > r^2} f_{\gamma',\tilde{\mathbf{Y}}^n|\tilde{\mathbf{X}}^n}(z^n|x_{w,j}^n) dz^n$$

$$+ \int_{\frac{1}{n}\|z^n - \tilde{\mathbf{H}}_\gamma^n x_{w,j}^n\|^2 < r^2} |f_{\gamma,\tilde{\mathbf{Y}}^n|\tilde{\mathbf{X}}^n}(z^n|x_{w,j}^n) - f_{\gamma',\tilde{\mathbf{Y}}^n|\tilde{\mathbf{X}}^n}(z^n|x_{w,j}^n)| dz^n \tag{333}$$

Note that from the triangular inequality, we have:

$$\left\| z^n - \tilde{\mathbf{H}}_\gamma^n x_{w,j}^n \right\| \tag{334}$$

$$\leq \left\| z^n - \tilde{\mathbf{H}}_{\gamma'}^n x_{w,j}^n \right\| + \left\| \left( \tilde{\mathbf{H}}_{\gamma'}^n - \tilde{\mathbf{H}}_\gamma^n \right) x_{w,j}^n \right\| \tag{335}$$

$$\leq \left\| z^n - \tilde{\mathbf{H}}_{\gamma'}^n x_{w,j}^n \right\| + r'\sqrt{n} \tag{336}$$

The last step follows from (89) and (90).

Therefore $\frac{1}{n}\left\| z^n - \tilde{\mathbf{H}}_\gamma^n x_{w,j}^n \right\|^2 > r^2$ implies:

$$\frac{1}{n}\left\| z^n - \tilde{\mathbf{H}}_{\gamma'}^n x_{w,j}^n \right\|^2 > (r - r')^2 \tag{337}$$

for

$$r > r' \tag{338}$$

This means that (333) is upper bounded by:

$$\int_{\frac{1}{n}\|z^n - \tilde{\mathbf{H}}_\gamma^n x_{w,j}^n\|^2 > r^2} f_{\gamma,\tilde{\mathbf{Y}}^n|\tilde{\mathbf{X}}^n}(z^n|x_{w,j}^n) dz^n$$

$$+ \int_{\frac{1}{n}\|z^n - \tilde{\mathbf{H}}_{\gamma'}^n x_{w,j}^n\|^2 > (r-r')^2} f_{\gamma',\tilde{\mathbf{Y}}^n|\tilde{\mathbf{X}}^n}(z^n|x_{w,j}^n) dz^n$$

$$+ \int_{\frac{1}{n}\|z^n - \tilde{\mathbf{H}}_\gamma^n x_{w,j}^n\|^2 < r^2} |f_{\gamma,\tilde{\mathbf{Y}}^n|\tilde{\mathbf{X}}^n}(z^n|x_{w,j}^n) - f_{\gamma',\tilde{\mathbf{Y}}^n|\tilde{\mathbf{X}}^n}(z^n|x_{w,j}^n)| dz^n \tag{339}$$

Hence, if $r > r'$ and

$$(r - r')^2 \geq N_E(1 + \varepsilon) \tag{340}$$

then (333) can be upper bounded by [32, (B2)]:

$$2e^{-n\alpha(\varepsilon)} + \int_{\frac{1}{n}\|z^n - \tilde{\mathbf{H}}_\gamma^n x_{w,j}^n\|^2 < r^2} |f_{\gamma,\tilde{\mathbf{Y}}^n|\tilde{\mathbf{X}}^n}(z^n|x_{w,j}^n) - f_{\gamma',\tilde{\mathbf{Y}}^n|\tilde{\mathbf{X}}^n}(z^n|x_{w,j}^n)| dz^n \tag{341}$$

The second term in (341) can be upper bounded by:

$$\int_{\frac{1}{n}\|z^n - \tilde{\mathbf{H}}_\gamma^n x_{w,j}^n\|^2 < r^2} |f_{\gamma,\tilde{\mathbf{Y}}^n|\tilde{\mathbf{X}}^n}(z^n|x_{w,j}^n) - f_{\gamma',\tilde{\mathbf{Y}}^n|\tilde{\mathbf{X}}^n}(z^n|x_{w,j}^n)| dz^n \tag{342}$$

$$= \int_{\frac{1}{n}\|z^n - \tilde{\mathbf{H}}_\gamma^n x_{w,j}^n\|^2 < r^2} f_{\gamma, \tilde{\mathbf{Y}}^n | \tilde{\mathbf{X}}^n}\left(z^n | x_{w,j}^n\right) |1 - \frac{f_{\gamma', \tilde{\mathbf{Y}}^n | \tilde{\mathbf{X}}^n}\left(z^n | x_{w,j}^n\right)}{f_{\gamma, \tilde{\mathbf{Y}}^n | \tilde{\mathbf{X}}^n}\left(z^n | x_{w,j}^n\right)}| dz^n \quad (343)$$

Recall that when $\frac{1}{n}\left\|z^n - \tilde{\mathbf{H}}_\gamma^n x_{w,j}^n\right\|^2 < r^2$, from (319)-(323) we have

$$1 - e^{ng(r,r')} \leq 1 - \frac{f_{\gamma', \tilde{\mathbf{Y}}^n | \tilde{\mathbf{X}}^n}\left(z^n | x_{w,j}^n\right)}{f_{\gamma, \tilde{\mathbf{Y}}^n | \tilde{\mathbf{X}}^n}\left(z^n | x_{w,j}^n\right)} \leq 1 - e^{-ng(r,r')} \quad (344)$$

Since $g(r, r') > 0$, we have

$$0 \leq |1 - \frac{f_{\gamma', \tilde{\mathbf{Y}}^n | \tilde{\mathbf{X}}^n}\left(z^n | x_{w,j}^n\right)}{f_{\gamma, \tilde{\mathbf{Y}}^n | \tilde{\mathbf{X}}^n}\left(z^n | x_{w,j}^n\right)}| \leq \max\{e^{ng(r,r')} - 1, 1 - e^{-ng(r,r')}\} \quad (345)$$

Note that $1 - e^{-x} \leq 1$ when $x \geq 0$. And there when $0 \leq x < 1$ $e^x - 1 \leq 2x$. Hence as long as

$$ng(r, r') < 1 \quad (346)$$

we have (343) upper bounded by:

$$\int_{\frac{1}{n}\|z^n - \tilde{\mathbf{H}}_\gamma^n x_{w,j}^n\|^2 < r^2} f_{\gamma, \tilde{\mathbf{Y}}^n | \tilde{\mathbf{X}}^n}\left(z^n | x_{w,j}^n\right) 2ng\left(r, r'\right) dz^n \quad (347)$$

$$\leq 2ng\left(r, r'\right) \quad (348)$$

Therefore as long as (338),(340) and (346) are satisfied, (331) is upper bounded by:

$$\int |f_{\gamma, \tilde{\mathbf{Y}}^n | \tilde{\mathbf{X}}^n}\left(z^n | x_{w,j}^n\right) - f_{\gamma', \tilde{\mathbf{Y}}^n | \tilde{\mathbf{X}}^n}\left(z^n | x_{w,j}^n\right)| dz^n$$

$$\leq 2e^{-n\alpha(\varepsilon)} + 2ng\left(r, r'\right) \quad (349)$$

Applying this result to (328), we have

$$2 \sum_w p_W(w) \int |f_{\gamma, \tilde{\mathbf{Y}}_G^n}\left(z^n\right) - f_{\gamma, \tilde{\mathbf{Y}}_{\mathcal{C}}^n | W}\left(z^n | w\right)| dz^n \quad (350)$$

$$\leq 4e^{-n\alpha(\varepsilon)} + 4ng\left(r, r'\right)$$

$$+ 2 \sum_w p_W(w) \int |f_{\gamma', \tilde{\mathbf{Y}}_G^n}\left(z^n\right) - f_{\gamma', \tilde{\mathbf{Y}}_{\mathcal{C}}^n | W}\left(z^n | w\right)| dz^n \quad (351)$$

and, we obtain (96) in Lemma 7.

Since $\gamma' \in S_M$, we can apply (78)-(79) to $\gamma'$, and bound (351) by:

$$4e^{-n\alpha(\varepsilon)} + 4ng\left(r, r'\right) + 12(2M + 1)^{2N_T N_E} \exp(-c'n) \quad (352)$$

Applying this result to (324)-(325), we obtain (97) in Lemma 7.

It remains to check that (338), (340) and (346) are satisfied. This is guaranteed by the definitions of $r$ and $r'$ in (90) and (91) and the condition (94) in the Lemma 7. Hence we have completed the proof of Lemma 7.

# APPENDIX G
## PROOF OF LEMMA 10

Consider two eavesdroppers, whose respective channel matrices are given below:

$$\tilde{\mathbf{H}}_\gamma = [\mathbf{I}, \mathbf{0}]\mathbf{U}_1 \qquad (353)$$

$$\tilde{\mathbf{H}}_{\gamma'} = [\mathbf{I}, \mathbf{0}]\mathbf{U}_2 \qquad (354)$$

Let $\mathcal{C}_1$ be any codebook from the ensemble $\{\mathcal{C}\}$ described in Section IV-C. Let $\mathcal{C}_2$ be

$$\mathcal{C}_2 = \mathbf{U}_2\mathbf{U}_1^{-1}\mathcal{C}_1 \qquad (355)$$

$$= \{\mathbf{U}_2\mathbf{U}_1^{-1}x^n, x^n \in \mathcal{C}_1\} \qquad (356)$$

Define the probability density function of a codebook, $f(\mathcal{C})$, as

$$f(\mathcal{C}) = \prod_{i,j} Q_{\tilde{\mathbf{X}}^n}(x_{i,j}^n) \qquad (357)$$

Since a unitary transform does not change the $L_2$ norm, we have

$$f(\mathcal{C}_1) = f(\mathcal{C}_2) \qquad (358)$$

We also observe from the maximum likelihood decoder (166), that the value of $\eta_\gamma$ for a given codebook $\mathcal{C}$, $\eta_\gamma(\mathcal{C})$, only depends on the set $\tilde{\mathbf{H}}_\gamma\mathcal{C}$, which is defined as:

$$\tilde{\mathbf{H}}_\gamma\mathcal{C} = \{\tilde{\mathbf{H}}_\gamma x^n : x^n \in \mathcal{C}\} \qquad (359)$$

Since

$$\tilde{\mathbf{H}}_\gamma\mathcal{C}_1 = \tilde{\mathbf{H}}_{\gamma'}\mathcal{C}_2 \qquad (360)$$

We have

$$\eta_\gamma(\mathcal{C}_1) = \eta_{\gamma'}(\mathcal{C}_2) \qquad (361)$$

Let $\mathbf{U}' = \mathbf{U}_2\mathbf{U}_1^{-1}$. Let $\mathcal{C}' = \mathbf{U}'\mathcal{C}$. Then we have

$$\eta_\gamma = \int \eta_\gamma(\mathcal{C}) f(\mathcal{C}) d\mathcal{C} \qquad (362)$$

$$= \int \eta_{\gamma'}(\mathbf{U}'\mathcal{C}) f(\mathbf{U}'\mathcal{C}) d\mathcal{C}, \qquad (363)$$

$$= \int \eta_{\gamma'}(\mathcal{C}') f(\mathcal{C}') d\mathcal{C}' \qquad (364)$$

$$= \eta_{\gamma'} \qquad (365)$$

Hence we have proved Lemma 10.

# APPENDIX H

## PROOF OF LEMMA 11

Let $\tilde{\mathbf{Y}}_i$ denote the part of $\tilde{\mathbf{Y}}_\gamma^n$ received during the $i$th channel use. We begin by proving:

$$\sum_{i=1}^{n} E\left[\|\tilde{\mathbf{Y}}_i\|^2\right] \leq nN_E(P+1) \tag{366}$$

We have

$$\sum_{i=1}^{n} E\left[\|\tilde{\mathbf{Y}}_i\|^2\right] \tag{367}$$

$$= \sum_{i=1}^{n} E\left[\|\tilde{\mathbf{H}}_\gamma \tilde{\mathbf{X}}_i + \tilde{\mathbf{N}}_i\|^2\right] \tag{368}$$

$$= \sum_{i=1}^{n} E\left[\|\tilde{\mathbf{H}}_\gamma \tilde{\mathbf{X}}_i\|^2\right] + E\left[\|\tilde{\mathbf{N}}_i\|^2\right] + 2\text{Re}\left[E\left[(\tilde{\mathbf{H}}_\gamma \tilde{\mathbf{X}}_i)^H \tilde{\mathbf{N}}_i\right]\right] \tag{369}$$

Since $\tilde{\mathbf{N}}_i$ is independent from $\tilde{\mathbf{X}}_i$ and $\tilde{\mathbf{N}}_i$ has 0 mean and unit variance, we find that (369) equals:

$$\sum_{i=1}^{n} E\left[\|\tilde{\mathbf{H}}_\gamma \tilde{\mathbf{X}}_i\|^2\right] + E\left[\|\tilde{\mathbf{N}}_i\|^2\right] + 2\text{Re}\left[E\left[(\tilde{\mathbf{H}}_\gamma \tilde{\mathbf{X}}_i)^H\right] E\left[\tilde{\mathbf{N}}_i\right]\right] \tag{370}$$

$$= \sum_{i=1}^{n} E\left[\|\tilde{\mathbf{H}}_\gamma \tilde{\mathbf{X}}_i\|^2\right] + E\left[\|\tilde{\mathbf{N}}_i\|^2\right] \tag{371}$$

$$= nN_E + \sum_{i=1}^{n} E\left[\|\tilde{\mathbf{H}}_\gamma \tilde{\mathbf{X}}_i\|^2\right] \tag{372}$$

Let $\tilde{\mathbf{H}}_j$ be the $j$th row of $\tilde{\mathbf{H}}_\gamma$. Then from (36), we have $\tilde{\mathbf{H}}_j^H \tilde{\mathbf{H}}_j = 1$. Using this result, we find that (372) equals:

$$nN_E + \sum_{i=1}^{n} \sum_{j=1}^{N_E} E\left[|\tilde{\mathbf{H}}_j \tilde{\mathbf{X}}_i|^2\right] \tag{373}$$

$$= nN_E + \sum_{i=1}^{n} \sum_{j=1}^{N_E} E\left[\tilde{\mathbf{X}}_i^H \tilde{\mathbf{H}}_j^H \tilde{\mathbf{H}}_j \tilde{\mathbf{X}}_i\right] \tag{374}$$

$$= nN_E + \sum_{i=1}^{n} \sum_{j=1}^{N_E} E\left[|\tilde{\mathbf{X}}_i|^2\right] \tag{375}$$

$$= nN_E + \sum_{j=1}^{N_E} \|\tilde{\mathbf{X}}^n\|^2 \tag{376}$$

$$\leq nN_E(1+P) \tag{377}$$

The last step follows from the following fact

$$\frac{1}{n}\|\tilde{\mathbf{X}}^n\|^2 \leq P, \quad \forall \tilde{\mathbf{X}}^n \in \mathcal{C} \tag{378}$$

Note that this is a stronger requirement than the average power constraint, in that it requires the power of *each* codeword not to exceed $P$. This is guaranteed by our codebook construction described in Section IV-C. Lemma 11 then follows by using the fact that the average power constrained random vector achieves the largest differential entropy when it has Gaussian distribution with i.i.d. components [37].

## APPENDIX I
## PROOF OF LEMMA 12

Let $B_{i_0,x^n}$ be the set of values of $\tilde{\mathbf{Y}}^n$ for which the decoder $\varphi_{\gamma'}$ outputs $x^n$ given the label $i_0$. Define $\eta_\gamma|x^n$ be the probability of decoding error for the eavesdropper indexed by $\gamma$ when the codeword $x^n$ is transmitted. Let $B_{x^n} = B_{i_0,x^n}$ with $i_0$ being the $i$ label of $x^n$. Let $r^2 = 2N_E$. Then we have

$$\eta_\gamma|x^n = \int_{z^n \notin B_{x^n}} f_\gamma(z^n|x^n)\, dz^n \tag{379}$$

$$\leq \int_{\|z^n - \tilde{\mathbf{H}}_\gamma x^n\|^2 \geq nr^2} f_\gamma(z^n|x^n)\, dz^n + \int_{\substack{\|z^n - \tilde{\mathbf{H}}_\gamma x^n\|^2 < nr^2 \\ z^n \notin B_{x^n}}} f_\gamma(z^n|x^n)\, dz^n \tag{380}$$

$$\leq e^{-n\alpha} + \int_{\substack{\|z^n - \tilde{\mathbf{H}}_\gamma x^n\|^2 < nr^2 \\ z^n \notin B_{x^n}}} f_\gamma(z^n|x^n)\, dz^n \tag{381}$$

Equation (381) follows from [32, (B2)]. Next, we apply (89) and (317)-(322) to the second term of (381) with $r^2 = 2N_E$ and $(r')^2 = \frac{2N_T N_E P}{M^2}$ and find that (381) is upper bounded by

$$e^{-n\alpha} + e^{ng(r,r')} \int_{\substack{\|z^n - \tilde{\mathbf{H}}_\gamma x^n\|^2 < nr^2 \\ z^n \notin B_{x^n}}} f_{\gamma'}(z^n|x^n)\, dz^n \tag{382}$$

$$\leq e^{-n\alpha} + e^{ng(r,r')} \int_{z^n \notin B_{x^n}} f_{\gamma'}(z^n|x^n)\, dz^n \tag{383}$$

Therefore

$$\eta_\gamma \leq e^{-n\alpha} + e^{ng(r,r')} \eta_{\gamma'} \tag{384}$$

$$\leq e^{-n\alpha} + 5((2M+1)^{2N_T N_E} + 1) e^{-(\bar{E}(\delta_n) - g(r,r'))n} \tag{385}$$

where the last step follows by applying (173). This concludes the proof of Lemma 12.